\documentclass[10pt]{iopart}
\usepackage{cite}  
\usepackage{bm}
\usepackage{footnote}
\usepackage{threeparttable}
\usepackage{graphicx}
\usepackage{float}
\usepackage{upgreek}
\usepackage{units}

\usepackage{color}

\begin{document}
\title[Formation mechanisms of striations in a filamentary DBD in atmospheric pressure argon]{Formation mechanisms of striations in a filamentary dielectric barrier discharge in atmospheric pressure argon}
\author{Aleksandar P. Jovanovi\' c$^{1*}$, Tom\'aš Hoder$^2$, Hans H\"oft$^{1}$, Detlef Loffhagen$^1$, Markus M. Becker$^1$}

\address{$^{1}$Leibniz Institute for Plasma Science and Technology (INP), Felix-Hausdorff-Str. 2, 17489 Greifswald, Germany}
\address{$^{2}$ Department of Physical Electronics, Faculty of Science, Masaryk University,  Kotlářská 2, 61137 Brno, Czech Republic}
\ead{$^*$aleksandar.jovanovic@inp-greifswald.de}
\vspace{10pt}
\begin{indented}
\item[]
\end{indented}

\today

\begin{abstract}
Formation mechanisms of striations along the discharge channel of a single-filament dielectric barrier discharge (DBD) in argon at atmospheric pressure are investigated by means of a time-dependent, spatially two-dimensional fluid-Poisson model. The model is applied to a one-sided DBD arrangement with a 1.5\,mm gap using a sinusoidal high voltage at the powered metal electrode. The discharge conditions are chosen to mimic experimental conditions for which striations have been observed. It is found that the striations form in both half-periods during the transient glow phase, which follows the streamer breakdown phase. The modelling results show that the distinct striated structures feature local spatial maxima and minima in charged and excited particle densities, which are more pronounced during the positive polarity. Their formation is explained by a repetitive stepwise ionisation of metastable argon atoms and ionisation of excimers, causing a disturbance of the spatial distribution of charge carriers along the discharge channel. The results emphasise the importance of excited states and stepwise ionisation processes on the formation of repetitive ionisation waves, eventually leading to striations along the discharge channel.
\end{abstract}

\vspace{2pc}

%
\maketitle
%

\ioptwocol

\section{Introduction}
\label{Intro}
Dielectric barrier discharges (DBDs) are broadly used as a source of non-thermal atmospheric-pressure plasmas in various technological applications, spanning from chemical processing and surface modification, over flow control to plasma medicine~\cite{Kogelschatz-2003-ID2019, Fridman-2005-ID2283, Corke-2010-ID3985, Brandenburg-2017-ID4908, Brandenburg-2018-ID5134}. 
Therefore, understanding the physical mechanisms required for their stable operation is of high interest. 
Experimental investigations have shown that under certain conditions, atmospheric-pressure plasmas undergo spontaneous mode transitions and form instabilities in the form of striations along the discharge channel~\cite{Staack-2008-ID2536, Schafer-2009-ID2558, Hoder-2011-ID2684, Zhu-2017-ID4243, Yamada-2018-ID5718}. 
Understanding the origin and formation mechanisms of these instabilities is not only intriguing from a physical point of view, but also relevant for stable operation of DBD applications.

The phenomenon of striations, i.e.\ the appearance of alternating bright and dark layers along the discharge channel, is well known for a long time~\cite{RaizerGDP, Robertson-1957-ID2990, Pekarek-1968-ID2988, Swain-1971-ID6089, Amemiya-1984-ID3761, Ammelt-1992-ID759}. 
Striated structures occur 
in various kinds of discharges at low~\cite{Robertson-1957-ID2990, Pekarek-1968-ID2988, Swain-1971-ID6089, Amemiya-1984-ID3761, Ammelt-1992-ID759, Godyak-2019-ID5324, Akishev-2022-ID6043} and (sub-)atmospheric pressure~\cite{Iza-2005-ID3575, Staack-2008-ID2536, Hoder-2011-ID2684, Zhu-2017-ID4243, Yamada-2018-ID5718, Yamada-2020-ID5717, Zhu-2020-ID6145}.
Besides experimental investigation of striations, usually performed by electrical or optical measurements~\cite{Swain-1971-ID6089, Amemiya-1984-ID3761, Ammelt-1992-ID759, Staack-2008-ID2536, Hoder-2011-ID2684, Zhu-2017-ID4243, Yamada-2018-ID5718, Godyak-2019-ID5324, Yamada-2020-ID5717, Zhu-2020-ID6145, Akishev-2022-ID6043}, theoretical investigations have been carried out as well~\cite{Golubovskii-1986-ID5278, Sigeneger-1997-ID6087, Sigeneger-1997-ID6088, Sigeneger-1998-ID1242, Golubovskii-2005-ID2252, Golubovskii-2011-ID2681, Golubovskii-2013-ID3336, Golubovskii2014, Kolobov-2006-ID2352,  Arslanbekov-2019-ID5981, Kolobov-2020-ID5437, Kolobov-2022-ID5861, Arslanbekov-2005-ID3742, Tahiyat-2022-ID5851, Boeuf-2022-ID5938, Iza-2005-ID3575, Hoder-2011-ID2684, Papadakis-2012-ID3620, Sigeneger-2016-ID3928, Sigeneger-2019-ID5343, Kawamura-2016-ID4210, Kawamura-2017-ID5997, Zhu-2020-ID6145}. 
For instance, the relaxation of electrons to spatially homogeneous state and different striation modes in collision dominated plasma normalised by pressure have been investigated using a kinetic approach~\cite{Sigeneger-1997-ID6087, Sigeneger-1997-ID6088, Sigeneger-1998-ID1242}. Besides this, the majority of theoretical studies have been carried out for low or moderate pressure~\cite{Golubovskii-1986-ID5278, Golubovskii-2005-ID2252, Golubovskii-2011-ID2681, Golubovskii-2013-ID3336, Golubovskii2014, Kolobov-2006-ID2352,  Arslanbekov-2019-ID5981, Kolobov-2020-ID5437, Kolobov-2022-ID5861, Arslanbekov-2005-ID3742, Tahiyat-2022-ID5851, Boeuf-2022-ID5938}, and fewer of them have dealt with discharges at atmospheric or sub-atmospheric pressure~\cite{Iza-2005-ID3575, Hoder-2011-ID2684, Papadakis-2012-ID3620, Sigeneger-2016-ID3928, Kawamura-2016-ID4210, Kawamura-2017-ID5997, Sigeneger-2019-ID5343, Zhu-2020-ID6145}. 
A thorough investigation of the stratification of discharges in inert gases at low pressure (ranging from 5\,mTorr to 2\,Torr) has been conducted using kinetic or discrete dynamic models~\cite{Golubovskii-1986-ID5278, Golubovskii-2005-ID2252, Golubovskii-2011-ID2681, Golubovskii-2013-ID3336, Golubovskii2014}.  
The formation of striations in low-current discharges at low pressure has been explained by the occurrence of integer and non-integer resonances caused by the non-local behaviour of electrons~\cite{Golubovskii2014}. 

Apart from the kinetic or discrete dynamic models, fluid and hybrid modelling studies of striations along the positive column in low-pressure discharges in argon and nitrogen have been reported in
~\cite{Kolobov-2006-ID2352, Arslanbekov-2019-ID5981, Kolobov-2020-ID5437, Kolobov-2022-ID5861}.  
Recently, the modelling of the appearance of striations in a nitrogen glow discharge at a pressure of tens of Torrs has been presented~\cite{Tahiyat-2022-ID5851}. Furthermore, particle-in-cell (PIC) and fluid models have been applied to describe their occurrence in low-pressure discharges in neon and argon~\cite{Boeuf-2022-ID5938}. 
Detailed reviews of 
experimental and numerical studies of striations in rarefied gases are given, e.g. in~\cite{Kolobov-2006-ID2352,Tsendin-2010-ID2649, Golubovskii-2011-ID2681, Golubovskii2014}.

In contrast to the extensive studies on striations in low-pressure discharges, the formation mechanisms of striations in discharges at atmospheric and sub-atmospheric pressure have been less thoroughly investigated to date, which can be attributed to the sporadic and often unstable occurrence of striated structures at (sub-)atmospheric pressure.
Moreover, previous works have shown that, depending on the discharge configuration, different mechanisms are responsible for the formation of striations. 
For example, striations in a plasma display panel cell with a Xe/Ne gas mixture at pressures in the range of 100--500\,Torr have been found to be caused by the combined effect of surface charges and non-local electron kinetics~\cite{Iza-2005-ID3575}.
In~\cite{Papadakis-2012-ID3620}, heating effects have been discussed as a possible reason of striations in a filamentary DBD in air at atmospheric pressure.
The analysis of striations in a single-filament DBD in pure argon by means of optical measurements and kinetic studies of the electrons based on the solution of the spatially inhomogeneous electron Boltzmann equation has led to the conclusion that they result from a spatial electron relaxation initiated by a local disturbance~\cite{Hoder-2011-ID2684}.
The same DBD arrangement as in~\cite{Hoder-2011-ID2684} has later been analysed by means of a spatially one-dimensional fluid model and the results have indicated that the interplay of direct ionisation and ionisation processes involving excited species could be a possible trigger for the striations observed in the experiment~\cite{Becker-2013-ID3200}.
The relevance of excited species and stepwise ionisation for perturbation of the ionisation budget, and thus for the formation of striations, has also been found in~\cite{Zhu-2020-ID6145,Yamada-2020-ID5717}.
In~\cite{Zhu-2020-ID6145}, step‐wise ionisation, deviation from Maxwellian
electron energy distribution and inhomogeneous gas heating has been found as causal for stratification of a single-filament discharge in argon with admixtures of helium and nitrogen at 150--300\,Torr.
The analysis of striations in an atmospheric pressure neon plasma jet by optical emission spectroscopy has revealed a modulated density of excited species along the discharge channel~\cite{Yamada-2020-ID5717}.
Furthermore, different non-linear dependences of ionisation and recombination rates on the electron density have been found to cause the appearance of striations in a filamentary argon plasma in an RF plasma jet at atmospheric pressure~\cite{Sigeneger-2016-ID3928,Sigeneger-2019-ID5343}.
The interplay of ionisation and recombination has also found to be an important factor in the striation of atmospheric-pressure RF and DC discharges in a He/H$_2$O gas mixture, where kinetic non-local effects have been found to induce an ionisation instability~\cite{Kawamura-2016-ID4210, Kawamura-2017-ID5997}.

These previously published results show that the governing mechanisms for the stratification of an atmospheric-pressure discharge are manifold and kinetic non-local effects do not always play a predominant role. 
In particular, further insights are needed into the role of excited species and the ionisation/recombination budget as triggers of ionisation instabilities.
Self-consistent fluid modelling has already proven to be a suitable tool for this purpose~\cite{Becker-2013-ID3200,Sigeneger-2016-ID3928,Sigeneger-2019-ID5343}.
To this end, the present article investigates striated structures in a single-filament DBD in pure argon at atmospheric-pressure by means of a well-established fluid-Poisson model. 
While the same model has been used in~\cite{Jovanovic-2021-ID5864,Jovanovic-2022-ID5985} to analyse the formation of the very first (non-striated) discharge channel and the streamer-surface interaction in the DBD, it is used here to study the particle gain and loss processes as well as the electron energy budget under periodic conditions.
In fact, this represents the first in-depth analysis of the formation mechanisms of striations in a filamentary DBD at atmospheric pressure.

The article is structured as follows. A brief description of the computational domain and the discharge arrangement is given in section~2. In the third section, an overview of the fluid model and solution procedure is provided. 
The modelling results are presented and discussed in section four. 
Finally, section 5 summarises the findings of the study.

\section{Discharge arrangement and computational domain}
\begin{figure}[!ht]
\centering
\includegraphics[scale=1]{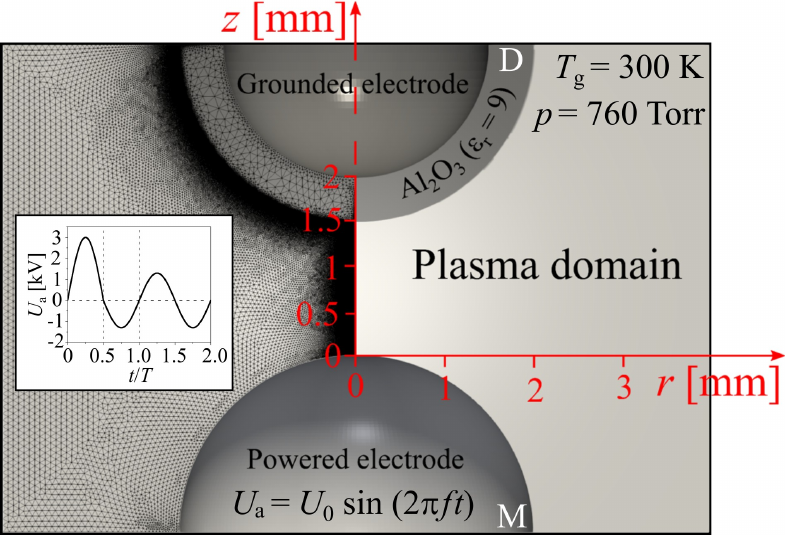}
\caption{%
Schematic representation of the discharge arrangement and illustration of the computational domains with the discretisation.}
\label{DomainIllust}
\end{figure}

The numerical 
analysis was performed for the asymmetric DBD configuration used by Hoder~\textit{et al.}~\cite{Hoder-2011-ID2684}, as illustrated in \fref{DomainIllust}.
This arrangement is frequently used to stabilise a single filament at the position with the shortest electrode distance, providing the stable conditions required for reliable measurements.
The configuration consists of two hemispherical stainless steel electrodes. 
The grounded electrode (denoted as D, dielectric) 
is covered by a 0.5\,mm thick dielectric layer (alumina, relative permittivity $\varepsilon_\mathrm{r} =9$). The powered electrode is bare (denoted as M, metal). 
The gas gap between the dielectric and the high-voltage metal electrode is $d=\unit[1.5]{mm}$. 
The DBD was driven by a sinusoidal applied voltage $U_\mathrm{a} = U_0\sin(2\pi f t)$ with the amplitude $U_\mathrm{0}=\unit[1.3]{kV}$ and frequency $f= \unit[60]{kHz}$ (corresponding to a period of ${T=\unit[16.67]{\upmu s}}$). 
The calculations were performed for atmospheric pressure ($p=\unit[760]{Torr}$) assuming a constant gas temperature of ${T_\mathrm{g}=\unit[300]{K}}$.
These parameters were chosen to be similar to the experimental conditions for which the stratification of the discharge channel has been observed in the experiment~\cite{Hoder-2011-ID2684}. 
Similar to the experiment, the discharge was initially ignited by applying an elevated high-voltage amplitude of $U_\mathrm{0}=\unit[3]{kV}$ in the first half-period, which was then reduced to the specified working voltage (as illustrated in the inset in~\fref{DomainIllust}). 
For the following analysis, the focus is on the subsequent periods after the ignition, i.e.\ the calculations were carried out until a quasi-periodic state was reached.

\section{Fluid-Poisson model}
\label{FluMod}

The same time-dependent, spatially two-dimensional fluid-Poisson model as in~\cite{Jovanovic-2021-ID5864,Jovanovic-2022-ID5985} was used in the present study.
It comprises the balance equations for the particle number densities of all considered species, the electron energy balance equation and Poisson's equation.
The set of balance equations 
\begin{equation}
\frac{\partial n_p}{\partial t} + \nabla \cdot \mathbf{\Gamma}_p=S_{p}
\label{eq:ContinuityEq}
\end{equation}
describe the spatiotemporal evolution of the particle densities $n_p$ of species $p$ (electrons, ions, neutral particles).
Here, $ \mathbf{\Gamma}_p$ denotes the particle flux and $S_{p}$ is the source term containing all gain and loss processes for the given species (either collisional or radiative)~\cite{Jovanovic-2021-ID5864}. 
The electric field $\mathbf{E} = - \nabla \phi$ is self-consistently determined by solving the Poisson equation
\begin{equation}
-\varepsilon_0 \varepsilon_\mathrm{r} \nabla^2 \phi = \rho
\label{eq:Poisson}
\end{equation} 
for the electric potential $\phi$, where $\rho=\sum_p q_{p} n_{p}$ is the space charge density, $q_p$ is the charge of species $p$ and $\varepsilon_0 \mathrm{\,and\,} \varepsilon_\mathrm{r} $ are the vacuum permittivity and the relative permittivity of the medium, respectively. 

The model operates within the framework of the local-mean-energy approximation, i.e.\ the electron rate and transport coefficients are incorporated as functions of the mean electron energy~\cite{Grubert-2009-ID2551}. 
To determine the mean electron energy  $u_\mathrm{e} = w_\mathrm{e}/n_\mathrm{e}$, the electron energy balance equation
\begin{equation}
\frac{\partial w_\mathrm{e}}{\partial t} + \nabla \cdot \mathbf{Q}_\mathrm{e} = -e_0 \mathbf{E} \cdot \mathbf{\Gamma}_\mathrm{e} +\widetilde{S}_\mathrm{e},
\label{eq:EEBE}
\end{equation}
is solved, where $w_\mathrm{e}$ and  $\mathbf{Q}_\mathrm{e}$ are the energy density and energy flux of electrons, $e_0$ denotes the elementary charge, $\widetilde{S}_\mathrm{e}$ represents the electron energy source term that takes into account the gain and loss of electron energy due to elastic and inelastic collisions, while the energy source term $-e_0 \mathbf{E} \cdot \mathbf{\Gamma}_\mathrm{e}$ represents the power input from the electric field.

The particle and energy fluxes are expressed using the 
drift-diffusion approximation 
\begin{equation}
\mathbf{\Gamma}_p = \mathrm{sgn}(q_p)\,b_p\,\mathbf{E}\, n_p - \nabla (D_p \, n_p),\\
\label{eq:DDP}
\end{equation}
\begin{equation}
\mathbf{Q}_\mathrm{e} = -\,\widetilde{b}_\mathrm{e}\,\mathbf{E}\, w_\mathrm{e} - \nabla (\widetilde{D}_\mathrm{e} \, w_\mathrm{e}),
\label{eq:DDQ}
\end{equation}
where $b_p$ and $D_p$ denote the mobility and the diffusion coefficient of the species $p$, while $\widetilde{b}_\mathrm{e} = 5b_\mathrm{e}/3$ and $\widetilde{D}_\mathrm{e}=5D_\mathrm{e}/3$ represent the mobility and the diffusion coefficient for the energy transport of electrons, respectively. 

To reduce the calculation time, which 
can be tremendous due to the need for multiple-period calculations, axial symmetry of the discharge is assumed.  
Accordingly, all equations are solved in cylindrical coordinates. 
Since the focus of this work is on the analysis of the axial stratification of the discharge channel (symmetry axis), this assumption appears to be reasonable.  
Note that this may affect the accuracy of the surface discharge description, as the assumption of axial symmetry is violated for off-axis discharge channels along the surface. 
However, the analysis of the formation mechanisms of striations along the filament between the hemispherical electrodes should not be affected.

The used reaction-kinetic model for argon involves the species $\mathrm{Ar}$, $\mathrm{Ar}^*$,  $\mathrm{Ar}_2^*$, $\mathrm{Ar}^+$, $\mathrm{Ar}_2^+$ and electrons~\cite{Sigeneger-2016-ID3928,Jovanovic-2022-ID5985}. 
Note that $\mathrm{Ar}^*$ and $\mathrm{Ar}_2^*$ represent lumped excited atomic and molecular (excimer) states, containing all higher excited levels.
The model takes into account elastic and inelastic electron collisions, such as excitation, de-excitation, direct and stepwise ionisation, and recombination. 
In addition, heavy-particle collisions, such as chemo-ionisation, neutral association and charge transfer, as well as radiative processes are taken into account. 
A detailed overview of the reaction scheme and  reaction rate coefficients is given in Appendix A. 
The rate and transport coefficients of electrons involved in equations \eref{eq:ContinuityEq} and~\eref{eq:EEBE}--\eref{eq:DDQ} were precalculated using the stationary, homogeneous electron Boltzmann equation in multi-term approximation~\cite{Leyh-1998-ID1222}. 
The mobilities of ionic species were taken from~\cite{Ellis-1976-ID3079} and the diffusion coefficients for these species were determined from the Einstein's relation. For $\mathrm{Ar}^*$, the diffusion coefficient was taken from~\cite{Phelps-1999-ID2920}, while no diffusion for $\mathrm{Ar}_2^*$ was assumed.

Equations~\eref{eq:ContinuityEq}--\eref{eq:EEBE} were closed by imposing a set of physically based boundary conditions and appropriate initial conditions.
In the case of the Poisson equation, Dirichlet boundary conditions were applied at the grounded ($\phi=0$) and at the powered  ($\phi=U_\mathrm{a}$) electrode, while the boundary condition $-\varepsilon_0 \varepsilon_\mathrm{r} \mathbf{E} \cdot \bm{\nu} = \sigma$ was used to account for the accumulated surface charges at the plasma-dielectric interface. 
The surface charge density $\sigma$ was determined by solving the balance equation
\begin{equation}
\frac{\partial \sigma}{\partial t} = \sum_p q_p\mathbf{\Gamma}_p \cdot \bm{\nu}
\end{equation}
at the plasma-dielectric interface and $\bm{\nu}$ denotes the outward normal vector to the boundary.
On the remaining boundaries, zero flux boundary conditions were imposed for the electric potential.
The boundary conditions applied for the balance equations at plasma-facing walls describe the partial reflection of particles and the ion-induced emission of secondary electrons, as further detailed in~\cite{Jovanovic-2021-ID5864}.
The same values for the reflection and secondary electron emission coefficients as in~\cite{Becker-2013-ID3200} were used. 
The reflection coefficient for neutral species was set to $0.3$ both for the dielectric and the metal surface, while for ions it was set to  $5 \times 10^{-4}$ (metal) and $5 \times 10^{-3}$ (dielectric), respectively. 
Reflection coefficients of $0.3$ (metal) and $0.7$ (dielectric) were used for electrons.
A secondary electron emission coefficient of $0.07$ at the metal and $0.02$ at the dielectric surface was used.
The mean energy of secondary electrons was set to $2$\,eV. 
In accordance with~\cite{Becker-2013-ID3200}, quasi-neutral initial conditions were employed by assuming a uniform density of $10^{12} \, \mathrm{m}^{-3}$ for heavy particles and $2 \times 10^{12} \, \mathrm{m}^{-3}$ for electrons. The initial mean electron energy was set to be $3$\,eV. No initial surface charges were considered on the dielectric surface.

The set of partial differential equations was solved in a fully coupled manner using the commercial software package COMSOL Multiphysics\textsuperscript{\textregistered}~\cite{ComsolMultiphysics}. The Matlab-Comsol toolbox for plasma models, MCPlas, was used for automated implementation of the model~\cite{Jovanovic-2021-ID5864}.
The calculations were performed on a compute server with two Intel Xeon E5-2690 CPUs, having in total 128\,GB of RAM. 
With a mesh consisting of 500\,000 elements and about 3 million degrees of freedom, calculations lasted on average 7 days per period (after reaching quasi-periodic state). 
A more detailed description of the model, its implementation, numerical details and the solution procedure can be found in~\cite{Jovanovic-2021-ID5864}. 

\section{Results and discussion}
\label{ResNDisc}
The discharge was initiated by the increased voltage amplitude, which leads to a strong discharge in the first half-period. 
The features of this particular discharge have been investigated recently~\cite{Jovanovic-2022-ID5985}. 
After the occurrence of the first discharge event, 
two periods of the applied sinusoidal voltage were required to establish a quasi-periodic state.

\begin{figure}[!ht]
\centering
\includegraphics[width = 8cm]{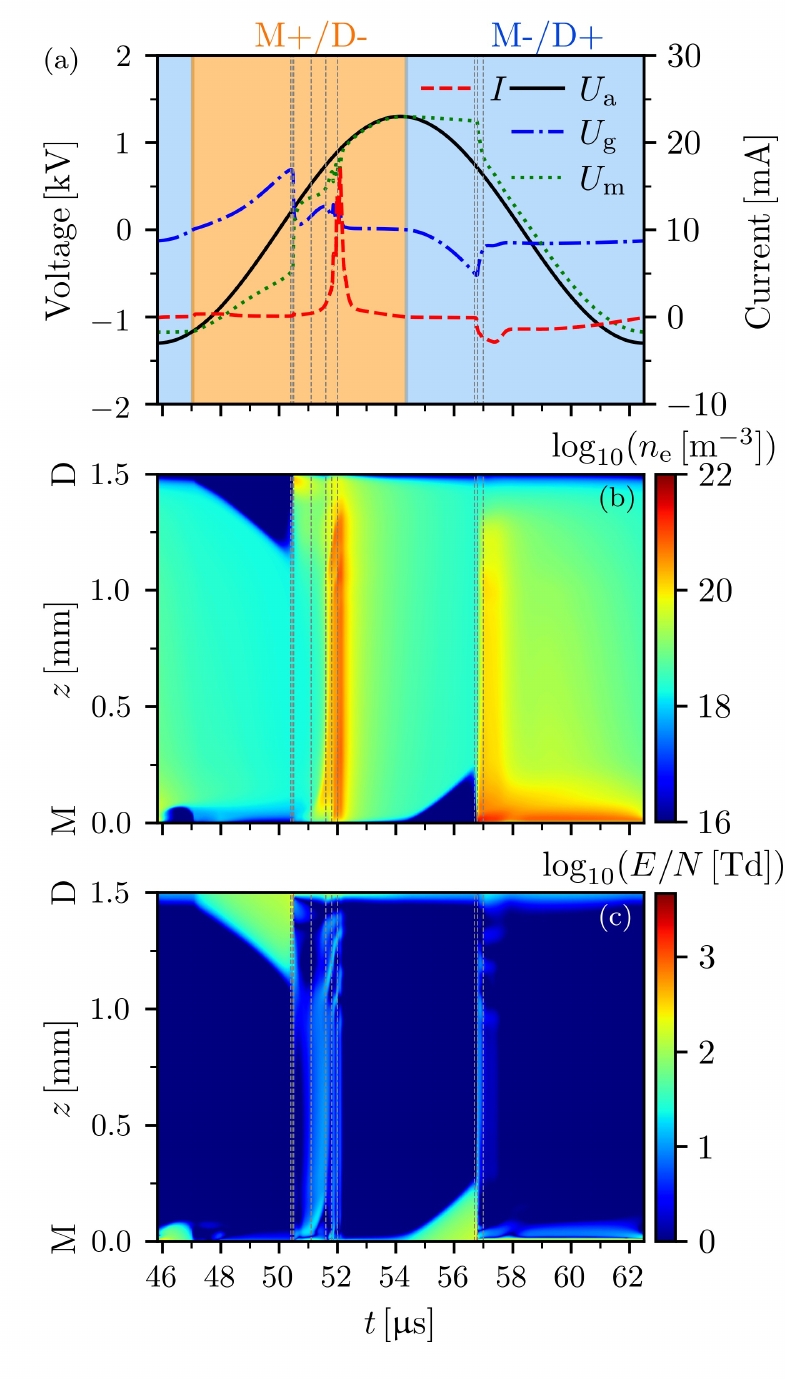}
\caption{Temporal evolution of the electric current $I$ and voltages $U_\mathrm{a}$, $U_\mathrm{g}$ and $U_\mathrm{m}$ (a); spatiotemporal evolution of the electron density (b) and reduced electric field (c) along the symmetry axis in log scale. 
One period after reaching the quasi-periodic state is shown.
`M' and `D' denote the metal electrode and the dielectric, respectively, while `$+$' and `$-$' indicate the momentary anode and cathode.
Dashed grey lines mark the characteristic times discussed in the manuscript.}
\label{ne2Dt}
\end{figure}

To illustrate the discharge behaviour after reaching a quasi-periodic state,~\fref{ne2Dt} shows the electric current, the applied voltage $U_\mathrm{a}$, the gap voltage $U_\mathrm{g}$ and the memory voltage $U_\mathrm{m} = U_\mathrm{a}-U_\mathrm{g}$ together
with the spatiotemporal evolution of the electron density and the reduced electric field $E/N$,
with $N$ being the background gas density, on the symmetry axis. 
Note that the gap voltage was determined as the potential difference between the dielectric and metal electrode tip and the electric current was calculated as the surface integral of the total current density at the metal electrode.
The given time interval ranges from the last quarter of the second to the last quarter of the third period and it can be seen that one discharge event occurs during each half-period. 

The first discharge occurs during the rising slope of the applied voltage in the positive half-period (\mbox{M$+$/D$-$,} i.e.\ the metal electrode is the momentary anode and the covered electrode is the momentary cathode), when the gap voltage reaches the breakdown voltage of about $\unit[700]{V}$. 
It manifests as a strong current peak with a maximum amplitude of approx.~$\unit[17]{mA}$~(cf.~\fref{ne2Dt}(a)). 
The second discharge starts during the falling slope of the applied voltage, at the moment when the gap voltage reaches the breakdown voltage of about $\unit[-520]{V}$. 
It lasts for most of the negative half-period (M$-$/D$+$, i.e.\ the covered electrode is the momentary anode and the metal electrode is the momentary cathode). 
The current is much weaker than during the previous discharge and reaches a peak value of about $\unit[3]{mA}$~(cf.~\fref{ne2Dt}(a)).

In line with the hypothesis of Černák~\textit{et al.}~\cite{Cernak-2020-ID5490},
the full development of the discharge is preceded
by a microscopic positive streamer impacting the
cathode. 
The generation of the positive streamer results from few microseconds long electron avalanching process, i.e.\ it has a multi-avalanche nature.  
Such mechanism takes place in both polarities. 
The impact of the microscopic positive streamer onto the cathode results in a steep current rise with a characteristic local maximum at the moment when the streamer reaches the surface, similar as observed in~\cite{Odrobina-1992-ID5473, Odrobina-1995-ID5487, Synek-2018-ID5102}. 
This phase is followed by a characteristic secondary ionisation wave, observed as an additional increase in the electron density in both half-periods (cf.~\fref{ne2Dt}(b)).
Although all characteristic phases in the DBD evolution, i.e.\ Townsend pre-phase, streamer phase, glow phase with cathode layer formation, and decay phase~\cite{Braun-1992-ID749, Steinle-1999-ID1363, Gibalov-2012-ID2757} can be distinguished (cf.~\fref{ne2Dt}(b) and (c)).
This observed behaviour generally agrees with the results of optical and electrical measurements for the low-current-mode discharges exhibiting a stratified discharge channel in the corresponding experiment~\cite{Hoder-2011-ID2684}.
This applies to both the magnitude and temporal duration of the discharge current and the spatiotemporal evolution of the discharge.
However, it should be noted that especially in the M$+$/D$-$ phase, the maximum value of the current is overestimated by the model due to the assumed axial symmetry of the discharge.
In fact, this assumption breaks as soon as surface discharge channels spread on the hemispherical electrodes during the transient glow phase~\cite{Jovanovic-2022-ID5985}.

The most pronounced phases with the highest electron density are the streamer and transient glow phases, which occur in both half-periods and have a strong influence on the striation formation. 
In the following, they are discussed separately for both half-periods.

\subsection{Discharge in the positive half-period (M$+$/D$-$)}
\begin{figure}[!t]
\centering
\includegraphics[scale=1]{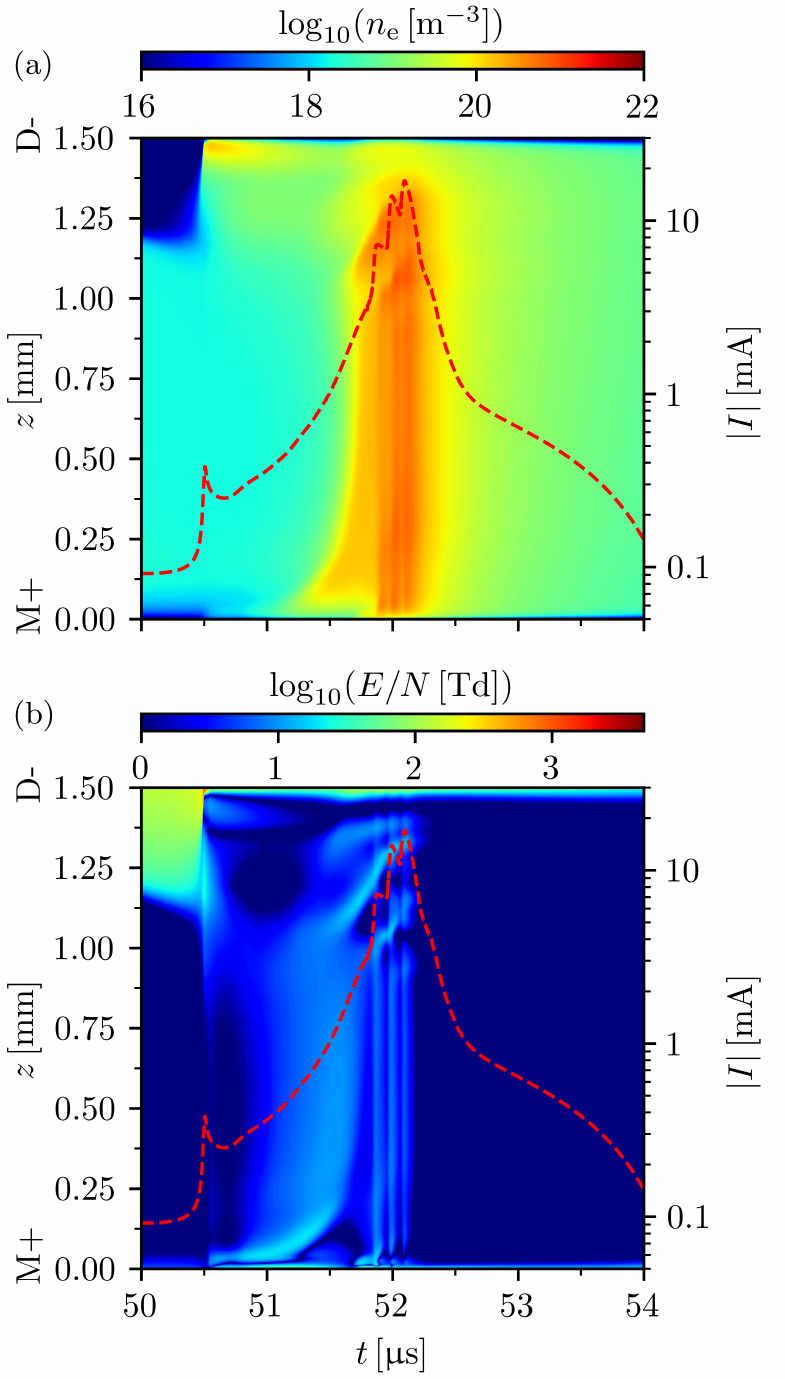}
\caption{Spatiotemporal evolution of the electron density (a) and reduced electric field $E/N$ (b) during the positive half-period. The current is overlaid over the data to illustrate the different phases in the discharge development. Due to strong variation, the decade logarithm of these quantities is displayed. }
\label{2Dt-PHP}
\end{figure}
To illustrate the discharge occurring during the positive half-period, the spatiotemporal evolution of the electron density and  reduced electric field $E/N$ are displayed in \fref{2Dt-PHP} along with the temporal evolution of the electric current. 
The streamer phase starts at $t = 50.40 $\,$\upmu$s and goes along with a rapid increase of the discharge current. 
The fast current increase is a result of the exponential increase of the electron density 
during the positive streamer propagation. 
As the streamer closes the gap, the current features a local maximum at  $\unit[50.50]{\upmu s}$ with a peak value of $\unit[0.36]{mA}$, similar to the observations in~\cite{Odrobina-1992-ID5473, Odrobina-1995-ID5487, Synek-2018-ID5102}. 
At the same time, the cathode layer is formed with a maximum reduced electric field of about $\unit[1000]{Td}$ in the thin sheath region (cf.~\fref{2Dt-PHP}(b)). 
 
The transient glow phase occurs afterwards and lasts until $t = \unit[54]{\upmu s}$.
During this phase (approximately $\unit[0.5]{\upmu s}$ after streamer arrival at the cathode), the electron density starts to increase near the momentary anode ($z = 0$).
From this point, the maximum of the electron density moves towards the cathode in the form of an ionisation wave (cf.~\fref{2Dt-PHP}(a)), as can also be seen in the spatiotemporal development of the reduced electric field (cf.~\fref{2Dt-PHP}(b)).
This process repeats over time, resulting in characteristic temporal structures of the electron density and the electric field as well as local peaks in the electric current (cf.~\fref{2Dt-PHP}(a) and (b)). 
At the same time, multi-peak structures appear along the discharge channel that resemble the striations observed in the experiment~\cite{Hoder-2011-ID2684}.
With additional charge production during this phase the electric current further increases to a maximum peak value of~$\unit[17]{mA}$. 

To better understand the mechanisms governing the streamer phase and the transient glow phase, both phases are discussed separately in the following. 

\subsubsection{Streamer phase}
\begin{figure}[t]%
	\centering%
	\includegraphics[scale=1]{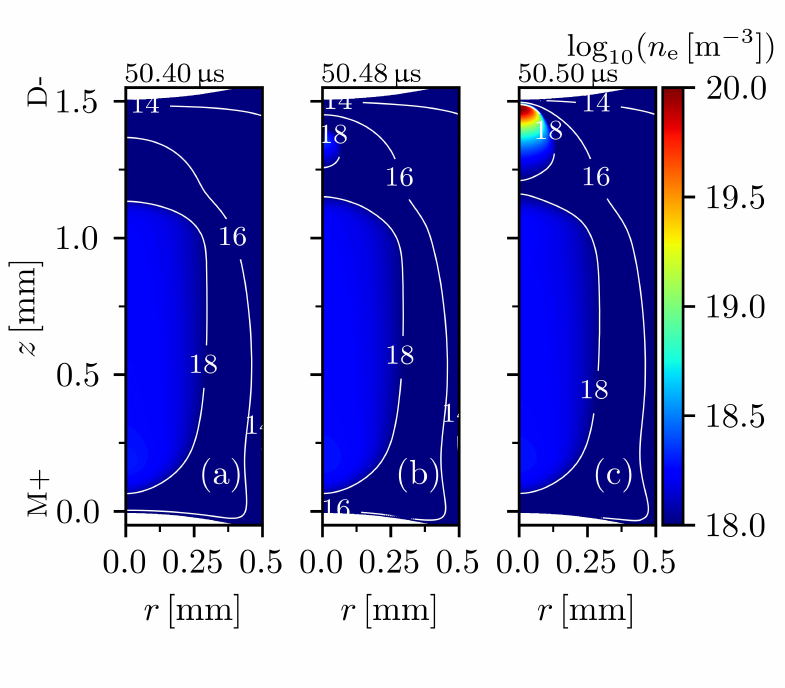}%
	\caption{Spatial distribution of the electron density in log scale during the streamer  phase at times  $t=\unit[50.40]{\upmu s}$ (a), $t=\unit[50.48]{\upmu s}$ (b) and $t=\unit[50.50]{\upmu s}$ (c). 
	}%
	\label{number_densities_2D-P1}%
\end{figure}

To illustrate the discharge development over the course of the streamer phase, the temporal evolution of the electron density  
is presented in~\fref{number_densities_2D-P1}.
At the beginning of the streamer phase, the discharge channel extends from approx.~$\unit[0.35]{mm}$ in front of the momentary cathode to less than\ $\unit[0.1]{mm}$ in front of the momentary anode and has a radius of about $\unit[0.25]{mm}$  (\fref{number_densities_2D-P1}(a)). 
Due to the increase of the electric field strength (cf.~\fref{2Dt-PHP}(b)), a local maximum of the electron density emerges between the bulk plasma and the momentary cathode (\fref{number_densities_2D-P1}(b)). 
This indicates a further space charge accumulation and local disturbance of the electric field.
The local field disturbance results in
the inception of a positive streamer propagating in the short gap between the bulk plasma acting as virtual anode and the dielectric.
From the movement of the ionising front (determined as the position of the local electric field maximum at the streamer head), the velocity of the streamer was estimated. 
It reaches a maximum value of around~$\unit[5 \times 10^{3}]{m/s}$ and has an average of $3 \times \unit[10^{3}]{m/s}$. This is in agreement with the starting velocity of the streamer bridging the entire gap during the initial discharge event~\cite{Jovanovic-2022-ID5985}, i.e.\ the propagation distance is too short to reach velocities typical for streamers in longer gaps.
When the streamer reaches the cathode (\fref{number_densities_2D-P1}(c)), the volume propagation stops and the cathode layer is formed. 
The radius of the streamer at this moment is about~$\unit[0.1]{mm}$.
This value is in agreement with the streamer radius of about~$\unit[80]{\upmu m}$ determined as full-width at half-maximum of Abel-inverted profiles measured in a coplanar surface DBD in argon~\cite{Simek-2012-ID6010}.

It is worth to mention that the streamer stops its propagation approx.~$\unit[15]{\upmu m}$ in front of the momentary cathode.
This behaviour differs from the case of the very first discharge event without surface charges on the dielectric surface, which has been thoroughly investigated in~\cite{Jovanovic-2022-ID5985}.
When no surface charges are present, the primary volume streamer stops its axial propagation approx.~$\unit[40]{\upmu m}$ in front of the dielectric, is deflected, and continues its propagation in radial direction. 
At the same time, an additional discharge develops between the deflected streamer and the dielectric surface~\cite{Jovanovic-2022-ID5985}.
In the present case of quasi-periodic conditions, surface charges remaining from the previous discharge modify the configuration of the electric field in the sheath region between the streamer head and the dielectric surface and prevent the immediate deflection of the volume streamer.

\begin{figure}[!t]
\centering
\includegraphics[scale=1]{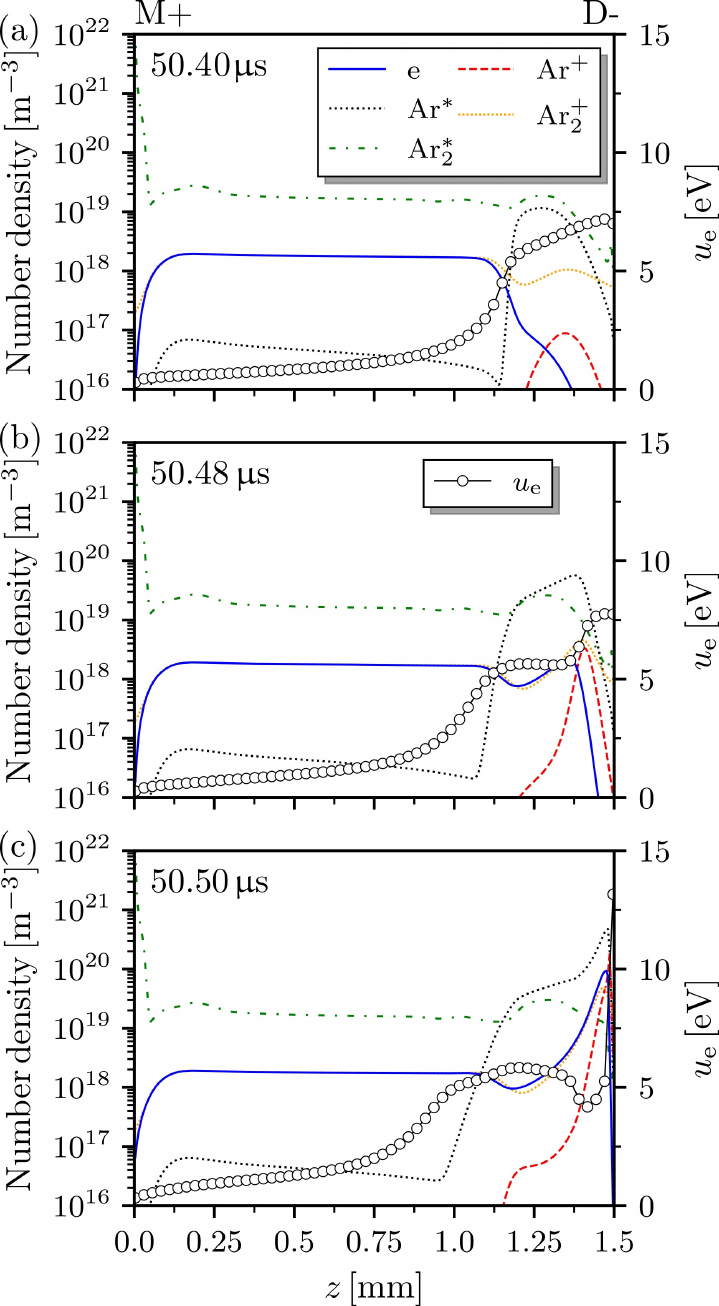}
\caption{Particle densities of all included species (left axis) and mean electron energy (right axis) along the symmetry axis ($r=0$) at the times  $t=\unit[50.40]{\upmu s}$ (a), $t=\unit[50.48]{\upmu s}$ (b) and $t=\unit[50.50]{\upmu s}$ (c).  Four regions can be distinguished, a cathode layer, negative glow, positive column and anode layer.}
\label{number_densities_axial_cut-P1}
\end{figure}

\begin{figure}[!t]
\centering
\includegraphics[scale=1]{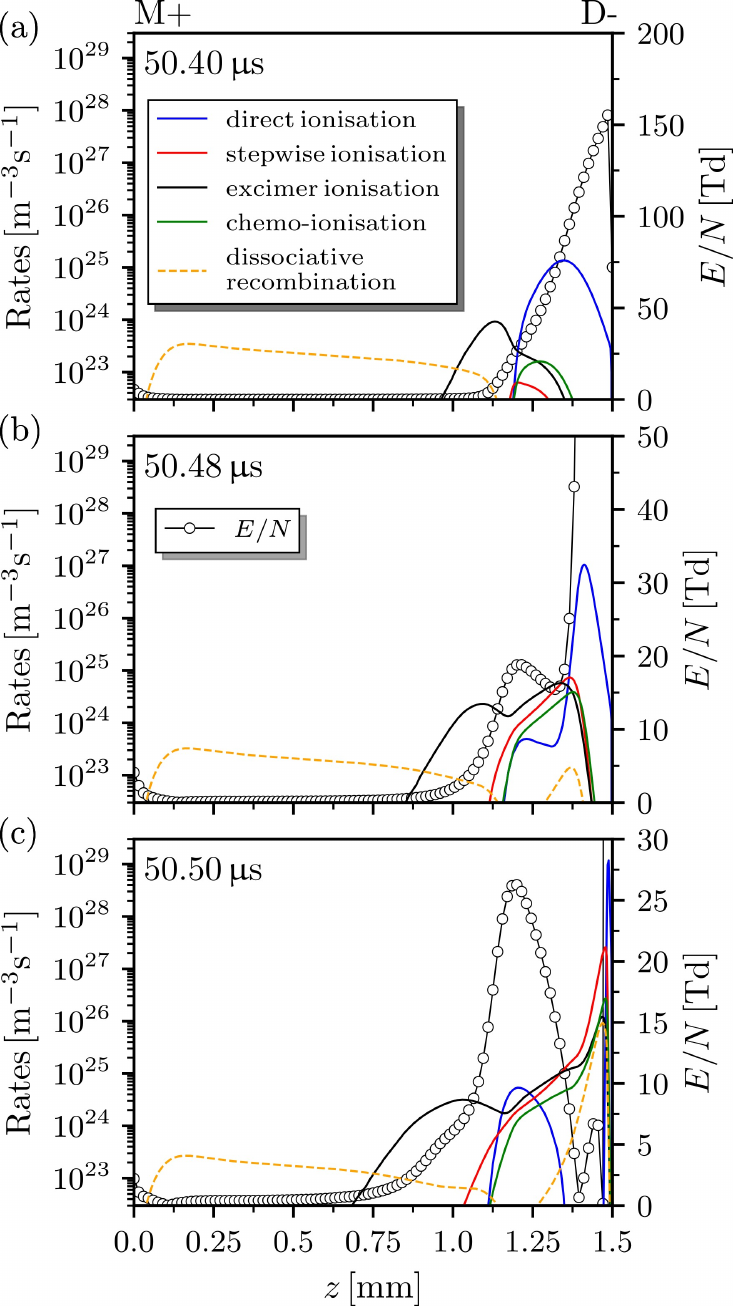}
\caption{Rates of electron production (left axis) and reduced electric field (right axis) along the symmetry axis at the times  $t=\unit[50.40]{\upmu s}$ (a), $t=\unit[50.48]{\upmu s}$ (b) and $t=\unit[50.50]{\upmu s}$ (c). The displayed range on $E/N$ axis is limited to emphasise the increase of the field behind the ionising front.}
\label{RatesP1}
\end{figure}

The axial profiles of the densities of electrons, ions and excited species as well as the mean electron energy during the streamer phase are presented in~\fref{number_densities_axial_cut-P1}.
In addition, the electron production rates and the reduced electric field are shown in \fref{RatesP1} to determine the dominant processes responsible for the streamer inception and propagation.
In~\fref{number_densities_axial_cut-P1}(a), it can be seen that at the moment of streamer inception, almost the whole gap is quasi-neutral except for the sheath regions 
near the surfaces.
Due to the strong atomic-to-molecular ion conversion, $\mathrm{Ar}_2^+$ is the dominant ion species at this instant.
Note that the densities of all species are uniform in the plasma bulk, i.e. there is no stratification. 

\begin{figure*}[!t]
\centering
\includegraphics[scale=1]{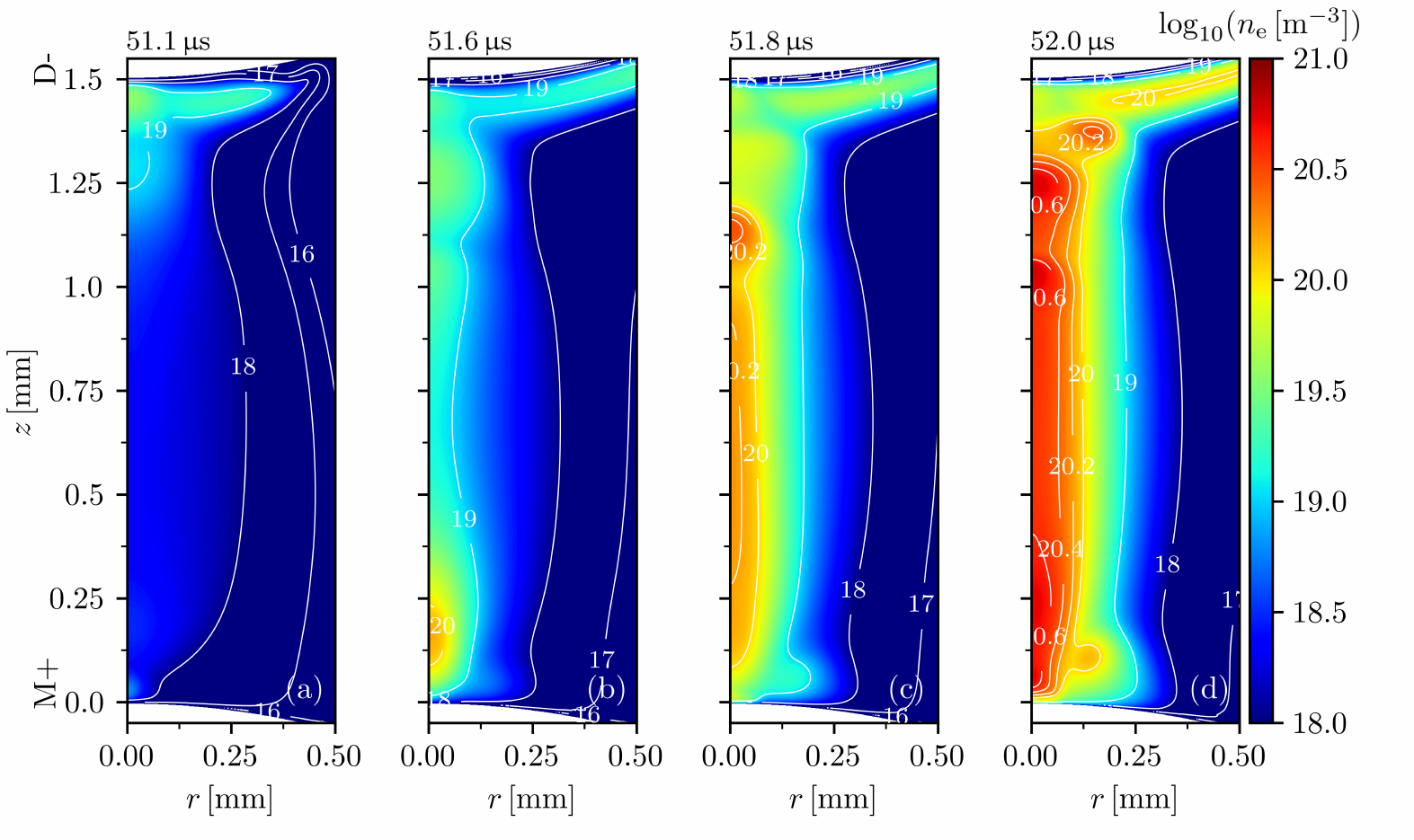}
\caption{Spatial distribution of the electron density in log scale during the transient glow  phase of the DBD at the times $t=\unit[51.1]{\upmu s}$ (a), $t=\unit[51.6]{\upmu s}$ (b), $t=\unit[51.8]{\upmu s}$ (c) and $t=\unit[52.0]{\upmu s}$ (d). 
}
\label{number_densities_2D-P2}
\end{figure*}

As the voltage increases, the drift of electrons towards the anode leads to an increase of the positive space charge in the sheath region near the momentary cathode~(\fref{number_densities_axial_cut-P1}(a)).
The increasing positive space charge and the accumulated negative charges on the dielectric (originating from the discharge in the previous half-period) enhance the electric field to about\ $\unit[155]{Td}$~(cf.~\fref{RatesP1}(a)) and, consequently, the mean electron energy rises to approx.~$\unit[7]{eV}$~(\fref{number_densities_axial_cut-P1}(a)) in the proximity of the momentary cathode. 
With the increase of the electric field, the influx of ions to the momentary cathode initiates the Townsend pre-phase. The electrons in this region now have an energy which is high enough to excite and ionise the ground state atoms (notice an increase in $\mathrm{Ar}^+$ ion density in~\fref{number_densities_axial_cut-P1}(a)), promoting the further increase of the space charge and the electric field. 
It should also be mentioned that after the direct ionisation, the electrons gain enough energy for excitation and additional ionisation of excimers in the sheath region. 
In the bulk plasma, the electron loss due to dissociative recombination is the dominant process, while electron production processes are negligible (\fref{RatesP1}(a)).

As soon as the electron density becomes high enough to significantly disturb the electric field, streamer inception occurs~\cite{RaizerGDP}.
The critical electron density here exceeds~$\unit[10^{18}]{m^{-3}}$.
As the positive streamer propagates towards the momentary cathode, the density of electrons and ions increases from about~$\unit[10^{18}]{m^{-3}}$ to approximately $\unit[10^{20}]{m^{-3}}$ (cf.~\fref{number_densities_2D-P1}(a) and \fref{number_densities_axial_cut-P1}(b) and (c)).
At the same time, the reduced electric field in front of the ionising front increases from $\unit[245]{Td}$ to about $\unit[1000]{Td}$. Behind the streamer head, $E/N$ drops quite fast from around $\unit[20]{Td}$ to less than $\unit[1]{Td}$ in the plasma bulk~(\fref{RatesP1}(b) and (c)). 
The maximum of the mean electron energy closely follows the movement of the ionising front. 
The mean electron energy increases from~$\unit[7.7]{eV}$ to the maximum of~$\unit[13.2]{eV}$ when the cathode layer is formed (\fref{number_densities_axial_cut-P1}(c)).  
Behind the ionising front, it is much lower (approx.~$\unit[5]{eV}$). 
After the streamer arrives at the cathode, it leaves an increased density of excited states and ions behind (\fref{number_densities_axial_cut-P1}(c)). 
Now, the spatial distribution of particle densities resembles a glow discharge (\fref{number_densities_axial_cut-P1}(c)~and~\fref{number_densities_2D-P1}(c)), with distinguishable cathode layer, negative glow, positive column and the anode layer, similar as in the experiment~\cite{Hoder-2011-ID2684}.

\begin{figure*}[!htb]
\centering
\includegraphics[scale=1]{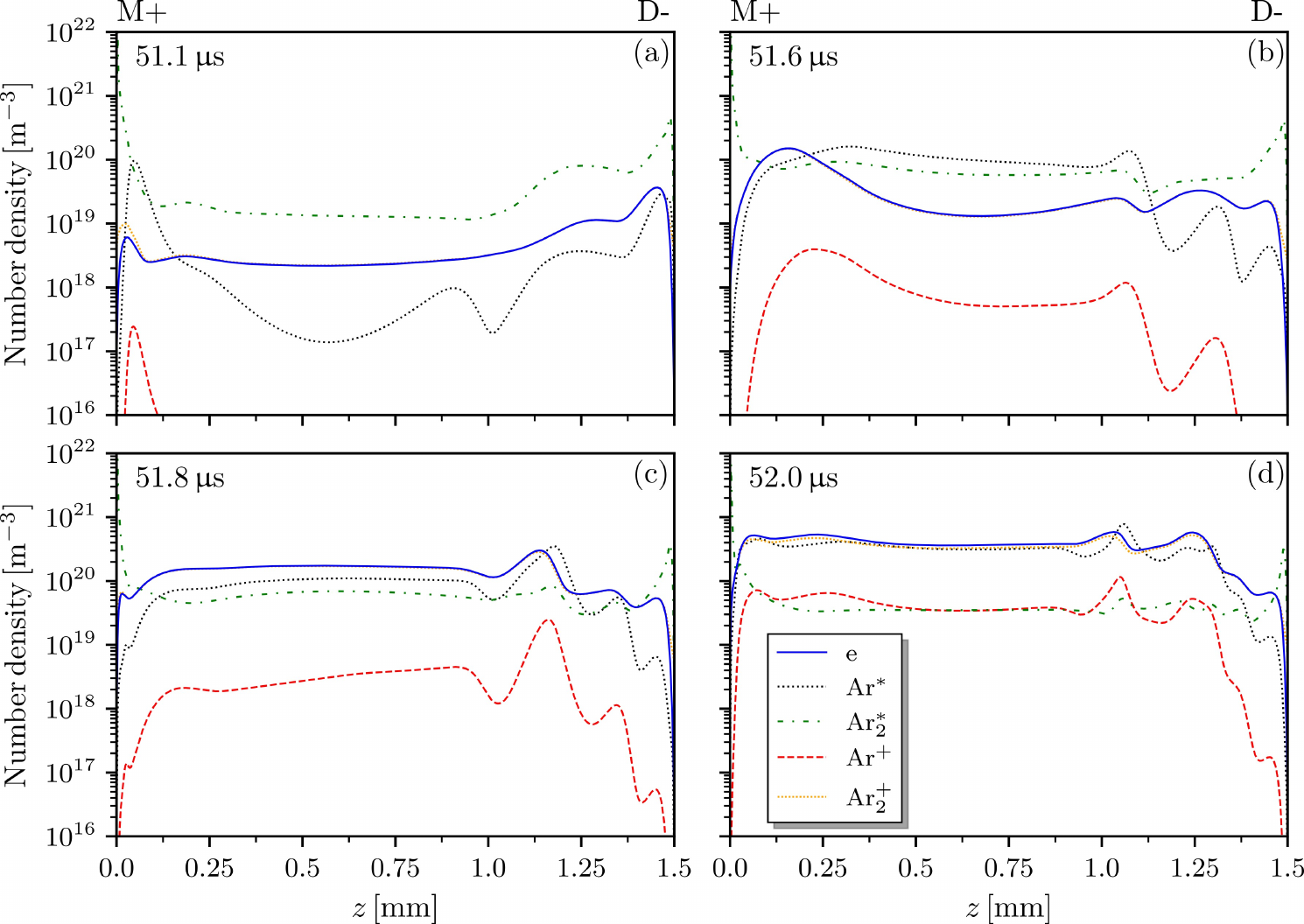}
\caption{Distribution of all particle densities along the symmetry axis ($r=0$) for the conditions of \fref{number_densities_2D-P2}.}
\label{number_densities_axial_cut2}
\end{figure*}

Although the impact of stepwise ionisation processes in the sheath region increases with time, their contribution is still orders of magnitude smaller than that of direct ionisation~(\fref{RatesP1}(b) and (c)). 
Thus, the direct impact ionisation is the main driving process for the streamer propagation towards the cathode. 
Behind the ionising front, the electrons gain enough energy to excite ground state atoms and ionise already excited states. 
This results in an additional local maximum in stepwise ionisation and ionisation of excimers (\fref{RatesP1}(b)).
In turn, the local space charge production due to these processes leads to a local increase of the reduced electric field to approx.~$\unit[26]{Td}$. 
The mean electron energy also increases, so that another local peak in the direct ionisation starts to emerge further from the cathode after the arrival of the streamer at the momentary cathode (cf.~\fref{RatesP1}(c)). 
These processes lead to the increase of the density of $\mathrm{Ar}^+$ and $\mathrm{Ar}^*$ behind the ionising front (\fref{number_densities_axial_cut-P1}(c)). 
The distribution of these species is crucial for the occurrence of the striated structures in the later phase of the discharge.

\subsubsection{Transient glow phase} 
The spatial distribution of the electron density 
is presented in~\fref{number_densities_2D-P2} at characteristic times during the transient glow  phase. 
It can be seen in \fref{number_densities_2D-P2}(a) that the surface discharge starts propagating just above the dielectric after the volume streamer has arrives at the cathode. 
The strong influx of secondary electrons produced during the surface discharge leads to a significant increase in the density of electrons and ions in the discharge channel. 
This results in the start of the transient glow  phase, which lasts much longer than the streamer phase and spans over several 
microseconds. 
It shows a more complex behaviour, eventually resulting in the stratification of the discharge channel.

\begin{figure*}[!htb]
\centering
\includegraphics[scale=1]{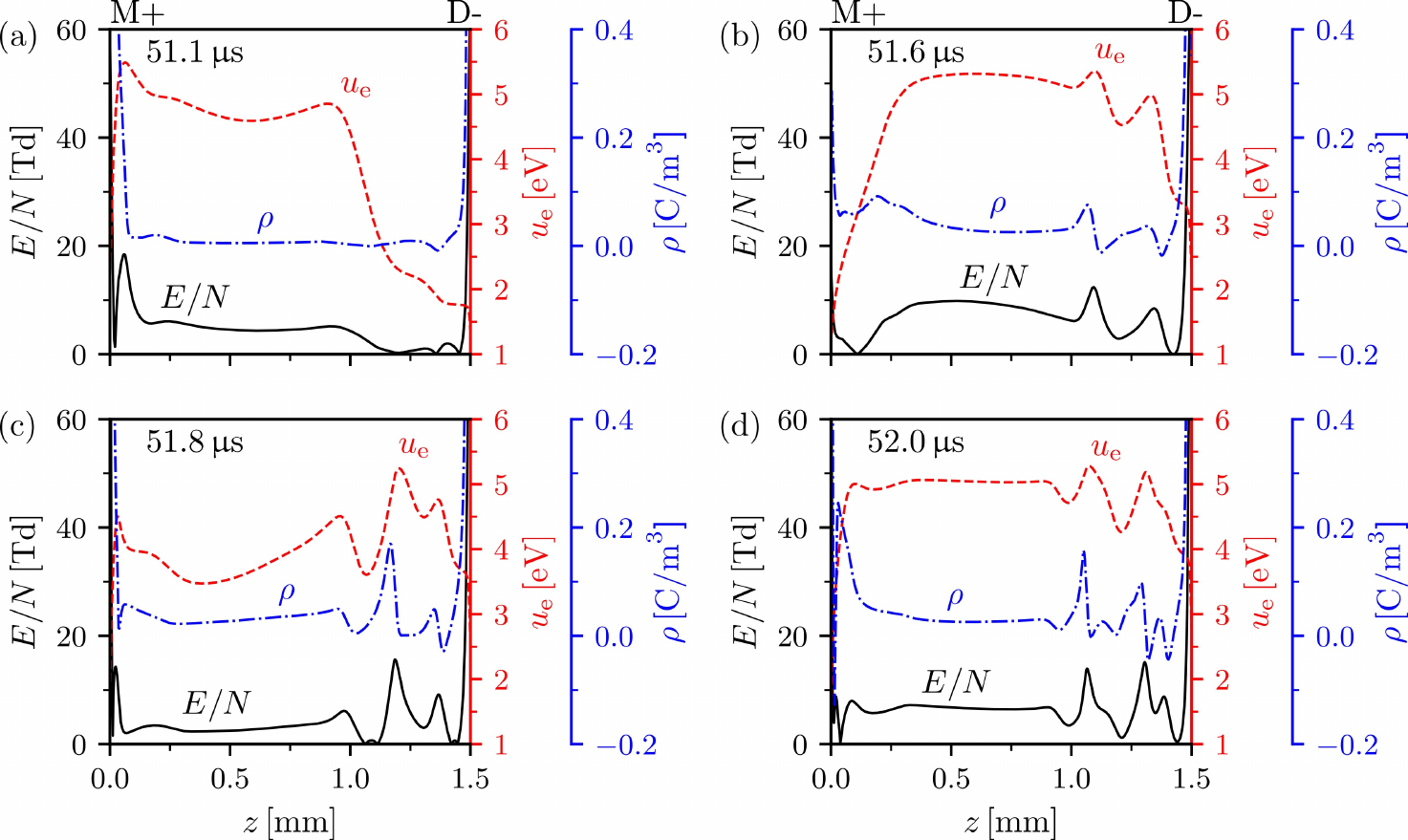}
\caption{Distribution of the reduced electric field $E/N$, mean electron energy $u_\mathrm{e}$ and space charge density $\rho$ along the symmetry axis ($r=0$) for the conditions of \fref{number_densities_2D-P2}. The displayed range is limited to emphasise the striations.}
\label{EUmRho}
\end{figure*}

\begin{figure*}[!hbt]
\centering
\includegraphics[scale=1]{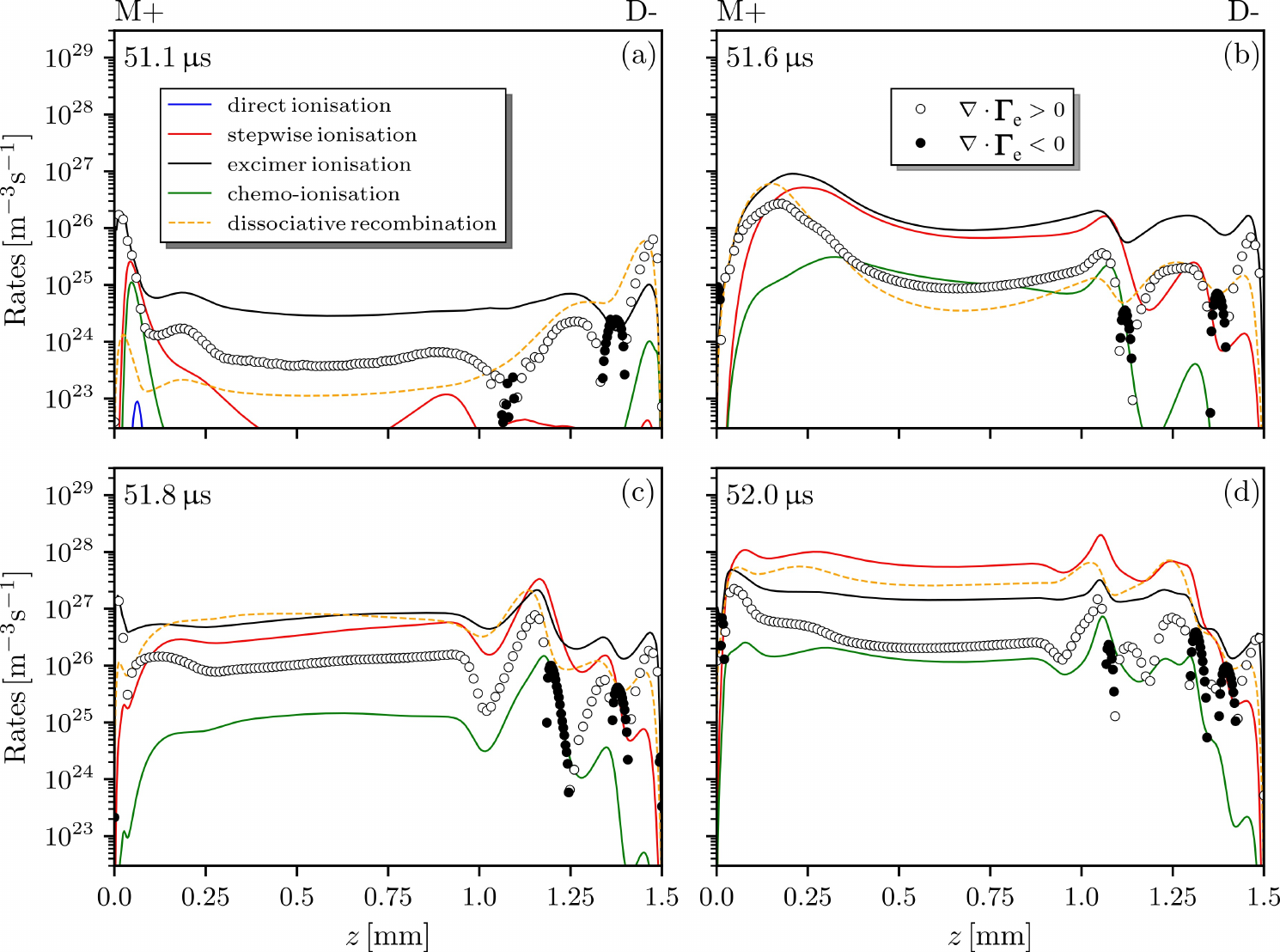}
\caption{Rates of electron production along the symmetry axis ($r=0$) for the conditions of \fref{number_densities_2D-P2}. The symbols represent the transport term, i.e. $\nabla \cdot \mathbf{\Gamma}_\mathrm{e}$, where gain and loss due to this term is distinguished.}
\label{RatesP2}
\end{figure*}

It can be seen that the influx of secondary electrons triggers the ionisation wave observed in~\fref{2Dt-PHP}(b). 
The ionisation wave moves from the anode towards the momentary cathode with a velocity of approx.~$\unit[100]{m/s}$, increasing the charge carrier densities uniformly in the gap and shifting the electron maximum towards the cathode~(cf.~\fref{number_densities_2D-P2}(c)). 
At this moment, a discharge channel with a radius of approx.~$\unit[120]{\upmu m}$ can be observed across the gap.
Local maxima of the excited states remaining from the streamer phase~(cf.~\fref{number_densities_2D-P1}(c)) cause the formation of distinct maxima of the electron density, which can be seen in~\fref{number_densities_2D-P2}(c)~and~(d). 
At the end of the transient glow  phase, four distinctive maxima can be observed. 
Note that an additional maximum develops in the surface part of the discharge channel~(cf.~\fref{number_densities_axial_cut-P1}(d)), similar to the experiment~\cite{Hoder-2011-ID2684}.

Figure~\ref{number_densities_axial_cut2} exhibits the particle densities of all species along the symmetry axis of the discharge during the transient glow  phase.  
It can be seen that at the beginning of this phase, the quasi-neutral channel extends almost from the cathode to the anode. 
Only in the narrow cathode and anode layer (less than~$\unit[100]{\upmu m}$ in front of cathode and anode), the density of $\mathrm{Ar}_2^+$ ions is higher than the density of electrons (\fref{number_densities_axial_cut2}(a)). 
Again, at the start of the transient glow  phase, the density of $\mathrm{Ar}^+$ ions has dropped due to their strong conversion into molecular ions. 
This positive space charge in the anode region results in a stronger electric field and higher mean electron energy, leading to an increase of the electron and $\mathrm{Ar}^+$ ion production~(\fref{number_densities_axial_cut2}(a)). 
As the secondary electrons produced during the surface discharge slowly drift towards the anode, they gain enough energy to further ionise the gas in the bulk plasma.
Consequently, a rather uniform increase of the charge carrier densities along the discharge channel can be observed (\fref{number_densities_axial_cut2}(b)). 
It should be stressed that the local maxima of the excited states that remained from the streamer phase lead to a localised enhancement in ionisation~(\fref{number_densities_axial_cut2}(c) and (d)). 
The ionisation wave ultimately leads to an increase of the density of charged species by almost two orders of magnitude in comparison to the streamer phase~(\fref{number_densities_axial_cut2}(d)). 

To better understand the formation of the striations, the spatial distributions of the reduced electric field $E/N$, mean electron energy $u_\mathrm{e}$ and space charge density $\rho$ (\fref{EUmRho}) as well as the main electron production rates (\fref{RatesP2}) are shown for the four characteristic times during the transient glow  phase.
At the start of the transient glow  phase, the field is weak in the plasma bulk (less than~$\unit[10]{Td}$), while it reaches approx.~$\unit[60]{Td}$ in the thin boundary layers near both surfaces~(\fref{EUmRho}(a)). 
Consequently, the electron energy gain from the electric field is small. 
This  leads to a much lower value of the mean electron energy of around~$\unit[5]{eV}$ in the bulk and~$\unit[5.5]{eV}$ near the anode in comparison to the streamer phase. 
Due to the lower mean electron energy, direct ionisation of the ground state atoms is no longer the dominant process.  
Instead, stepwise ionisation and the ionisation of the excimers is the main source of electrons at this moment (\fref{RatesP2}(a)). 
The slightly higher mean electron energy near the momentary anode leads to an increase in the density of atomic argon ions in the anode layer (cf.~\fref{number_densities_axial_cut2}(a)), while chemo-ionisation processes are negligible at this moment.

\begin{figure}[!htb]
\centering
\includegraphics[scale=1]{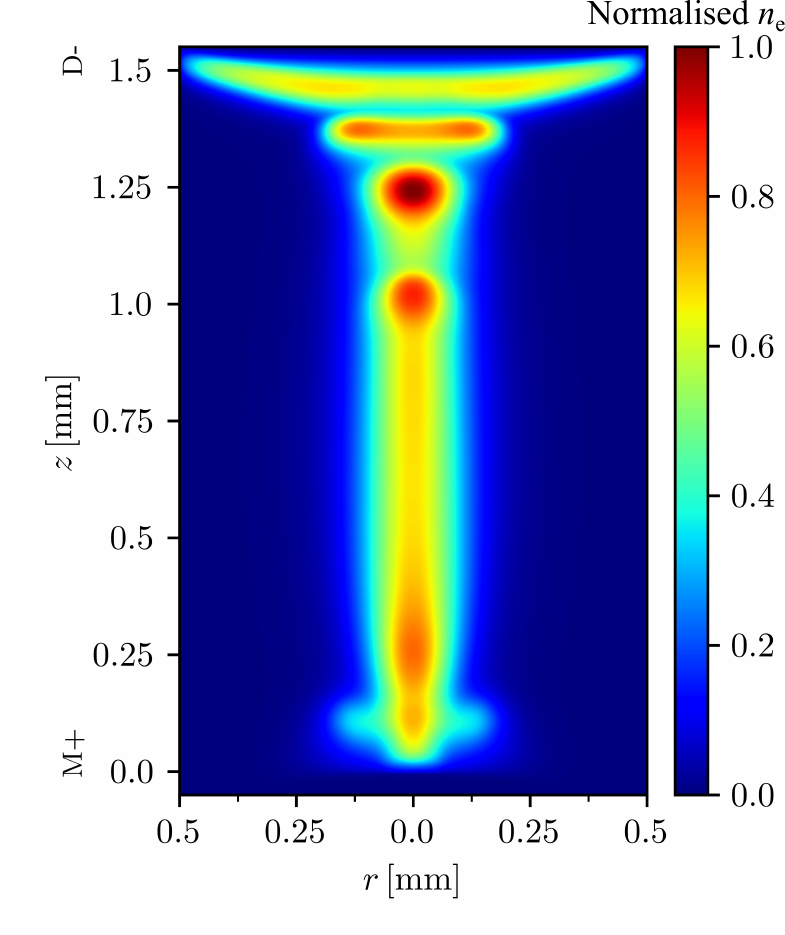}
\caption{Abel-transformed distribution of the electron density at the time of most intense striations ($t=\unit[52.0]{\upmu s}$).}
\label{StriationsAbel}
\end{figure} 

Although at the start of the transient glow  phase the ionisation is strongest in the sheath region near the momentary anode, the ionisation also occurs along the entire discharge channel due to the high density of metastable argon atoms and molecules in plasma bulk (cf.~\fref{number_densities_axial_cut2}(b)).
Note that the chemo-ionisation processes increase due to the increase in the Ar$^*$ density. However, their contribution is still orders of magnitude lower than that of stepwise ionisation and ionisation of excimers. 
Secondary electrons originating from the surface discharge now play an important role in the further discharge development. 
Namely, on their way to the anode, these electrons accumulate high enough energy 
for stepwise ionisation of excited states (\fref{RatesP2}(b)--(d)). 
The ionisation is enhanced at the locations where the excited states have a maximum. 
Depending on the position where the inception of the streamer in the streamer phase occurred, the maxima are closer or further away from the momentary cathode~(cf.~\fref{number_densities_axial_cut2}(c)). 
These local space charge maxima, in turn, lead to the strong modulation of the electric field with characteristic peaks along the discharge channel~(approx.~$\unit[15]{Td}$)~(\fref{EUmRho}(b)--(d)).  
The mean electron energy varies around~$\unit[5]{eV}$ along the gap, with distinctive peaks up to about~$\unit[5.5]{eV}$ and dips near the electric field maxima. 
The process repeats over time, producing distinct maxima in all three presented quantities. 
Note that subsequent faster ionising waves, observed as a characteristic temporal structure in the electron density and the electric field (cf.~\fref{2Dt-PHP}), originate from the surge of secondary electrons due to the mode change in the surface discharge. 
This repetitive behaviour provides the conditions for the formation of the observed striations along the axis~(cf.~\fref{number_densities_axial_cut2}(b)--(d)). 
It is important to mention that between field maxima, a field reversal is observed ($\rho < 0$ in~\fref{EUmRho}(b)--(d)), indicating that the electrons are trapped in these regions, further enhancing this effect. This is similar to the findings reported in~\cite{Godyak-2019-ID5324, Boeuf-2022-ID5938}. 

It should be stressed that the electron production processes are an order of magnitude higher than the loss of electrons due to their transport. 
Thus, the electron loss during this phase is reaction-dominated as shown in~\fref{RatesP2}(a)--(d). 
Note that the three-body recombination process is negligible, while the contribution of the dissociative electron-ion recombination process varies over time.
However, it is consistently lower than ionisation processes in the largest part of the discharge channel.
These findings are in agreement with previous works, where the importance of excited states, stepwise ionisation processes and the balance between ionisation and recombination processes for striation formation have been reported~\cite{Becker-2013-ID3200,Sigeneger-2016-ID3928,Sigeneger-2019-ID5343,Zhu-2020-ID6145}.

It is important to note that the intensity profiles obtained from optical measurements cannot be directly compared to the density profiles shown here. The present model includes lumped states for excited argon atoms and molecules and an extended reaction kinetics model is required for prediction of emission profiles~\cite{Stankov-2022-ID6138}.
Therefore, \fref{StriationsAbel} displays an artificial emission profile, which was calculated from the spatial distribution of the electron density after line-of-sight integration (Abel transformation was done using the pyAbel software~\cite{Hickstein-2019-ID5472}). 
Although the striations can be distinctively observed, the assumed axial symmetry leads to a broadening of the intensity maximum close to the momentary cathode. 
In the experiment, surface streamers follow a random direction, resulting in localised peaks close to and just above the surface~\cite{Hoder-2011-ID2684}.

\subsection{Discharge in the negative half-period (M$-$/D$+$)}
\begin{figure}[!htb]
	\centering
	\includegraphics[scale=1]{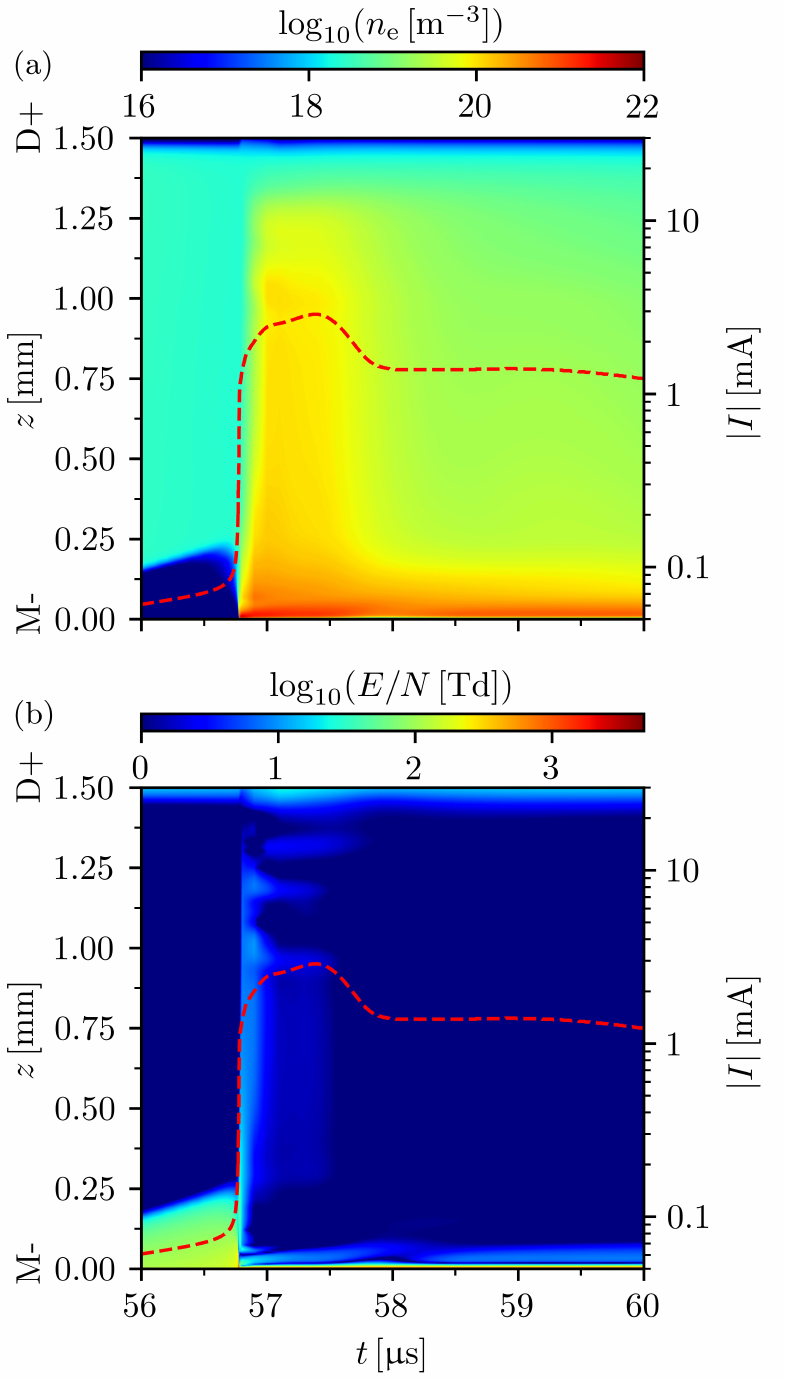}
	\caption{Spatiotemporal evolution of the electron density $n_\mathrm{e}$ (a) and reduced electric field $E/N$ (b) during the negative half-period. The current is placed over the data to illustrate the different phases of the discharge development. Due to strong variation, the decade logarithm of these quantities is displayed.}
	\label{2Dt-NHP}
\end{figure}

To illustrate the discharge during the negative half-period, where the metal electrode acts as cathode ($\mathrm{M}$-/$\mathrm{D}$+), the spatiotemporal evolution of the electron density and  the reduced electric field is displayed in \fref{2Dt-NHP} together with the electric current. 
It can be seen that the discharge evolution exhibits similar features as in the positive half-period.
The streamer propagation and transient glow  phases can be distinguished here as well. 
However, the transition between them is now continuous and much faster than in 
the positive half-period. 
The streamer propagation results in an increase of the electron density and electric field, causing the steep increase of the electric current. 
The characteristic local maximum is less pronounced and manifests as the change of the slope of the current. 
During the transient glow phase, the current increase is slower. 
The current reaches a maximal value of about~$\unit[3]{mA}$. 
Afterwards, it starts to slowly decay  during most of the negative half-period. 
The current amplitude is lower compared to the transient glow phase during the positive half-period, 
which is due to the absence of the surface discharge on the dielectric. 
Moreover, the lack of secondary electrons from the surface discharge prevents the temporal structure formation in the electron density and electric field during the transient glow phase.

\begin{figure}[!t]
\centering
\includegraphics[scale=1]{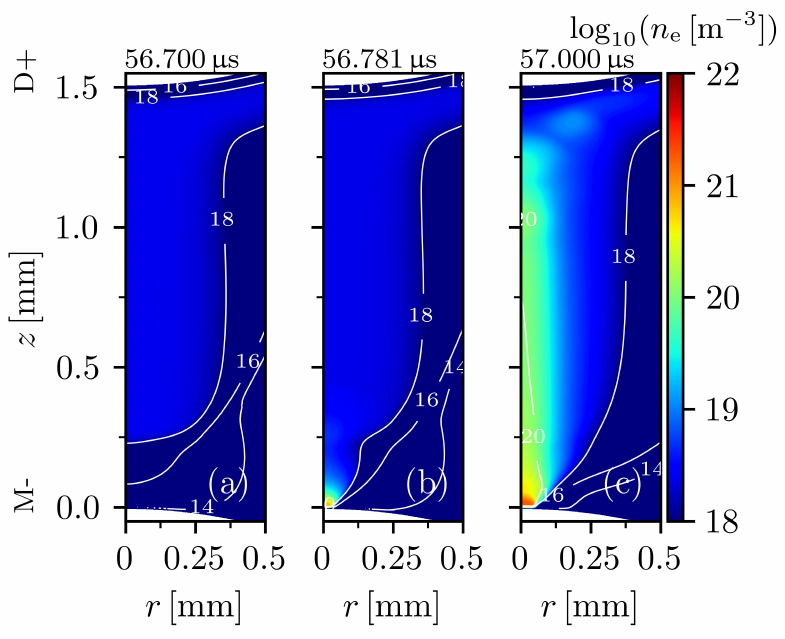}
\caption{Spatial distribution of the electron density in log scale during the discharge in the negative half-period at the times $t = 56.700 \upmu s$ (a), $t = 56.781 \upmu s$ (b) and $t = 57.000 \upmu s$ (c). 
}
\label{number_densities_2D-NHP}
\end{figure}

\Fref{number_densities_2D-NHP} illustrates the   development of the electron density during the negative half-period.
Slightly before the streamer inception takes place, a distinct discharge channel extends from approx.~$\unit[0.25]{mm}$ in front of the momentary cathode to the momentary anode.  
Note that at this time $t=\unit[56.700]{\upmu s}$, the channel still covers a part of the dielectric (\fref{number_densities_2D-NHP}(a)). 
With fulfilment of the streamer inception condition, the streamer starts to propagate towards the momentary cathode (cf.~\fref{2Dt-NHP}). 
It closes the gap in about $\unit[7]{ns}$ (\fref{number_densities_2D-NHP}(b)). 
From the movement of the ionising front (cf.~\fref{2Dt-NHP}(b)), the streamer propagation velocity was determined.
It is similar in magnitude to the streamer phase during the positive half-period ($\unit[4.5 \times 10^3]{m/s}$).

\begin{figure}[!t]
\centering
\includegraphics[scale=1]{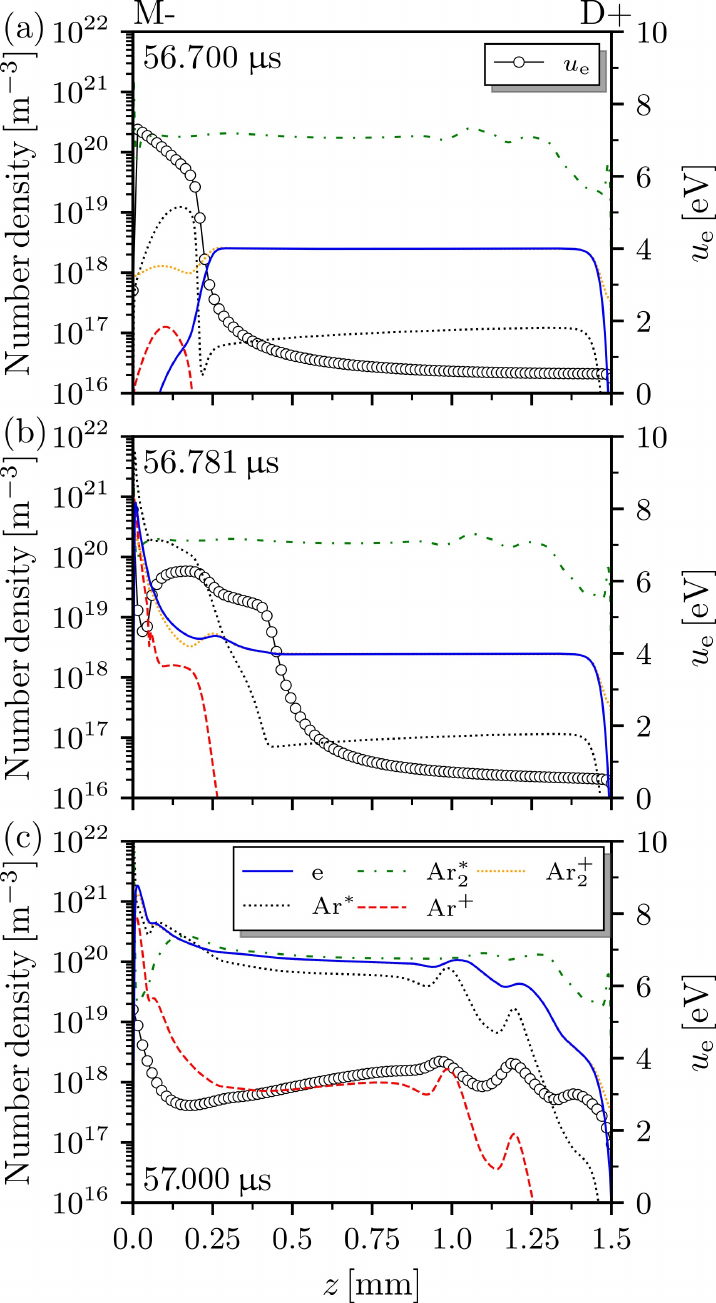}
\caption{Particle densities of all considered species (left axis) and mean electron energy (right axis) along the symmetry axis ($r=0$) for the conditions of \fref{number_densities_2D-NHP}.}
\label{DischargeNHP}
\end{figure}

\begin{figure}[!t]
\centering
\includegraphics[scale=1]{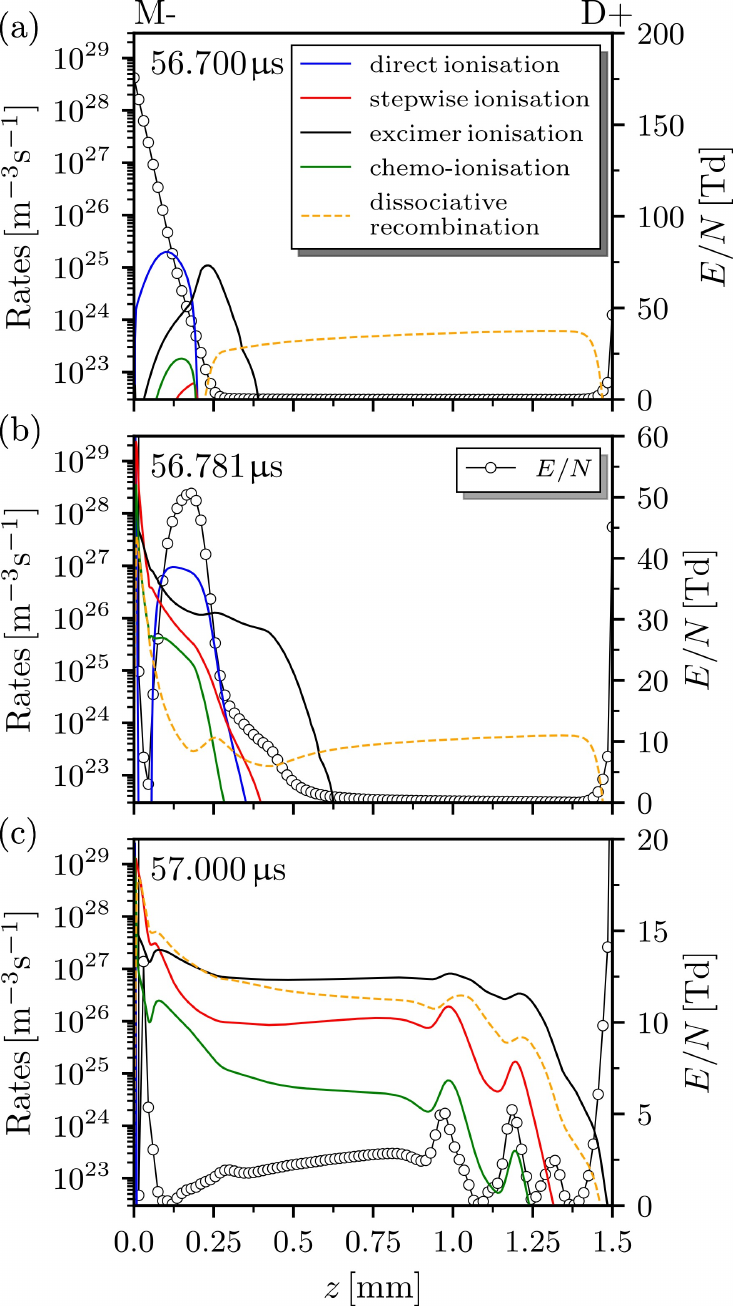}
\caption{Rates of electron production (left axis) and reduced electric field (right axis) along the symmetry axis ($r=0$) for the conditions of \ref{number_densities_2D-NHP}. 
The $E/N$ range is limited to emphasise the field modulation.}
\label{RatesNHP}
\end{figure}

In \fref{number_densities_2D-NHP}(b) it can be seen that the electron density reaches a maximum in the proximity of the momentary cathode at the end of the streamer phase. 
As the propagation slows down, a thin cathode layer forms~$\unit[5]{\upmu m}$ in front of the surface, resulting in a strong electric field. 
In contrast to the discharge in the positive half-period, the surface discharge along the cathode or radial expansion of the filament on the metal electrode are absent. 
This results in a much weaker electric current.
The discharge remains localised close to the symmetry axis with only narrow broadening near the tip of the momentary cathode (M$-$/D$+$). 
An increase of the electron density along the discharge channel, with less pronounced localised maxima in the proximity of the dielectric, is found during the transient glow phase~(\fref{number_densities_2D-NHP}(c)). 
Note that the maximum of the electron density remains at the tip of the momentary cathode over the whole time of the discharge duration (\fref{number_densities_2D-NHP}(b) and (c)). 
This maximum masks the striations, making them barely visible (cf.~\fref{number_densities_2D-NHP}(c)). 

To better understand the dominant processes during the discharge and to further analyse how they lead to the striation formation, the axial distribution of the particle densities of all species, the mean electron energy (\fref{DischargeNHP}), the electron production rates and the reduced electric field (\fref{RatesNHP}) are shown at radial position $r=0$. 
Just before the streamer inception (\fref{DischargeNHP}(a)), the densities of all species, except argon excimers, are uniform in the plasma bulk. 
This indicates that the spatial distribution of charge carriers left over from the discharge in the previous half-period does not play a role in the formation of the striations. 
Most of the gap is quasi-neutral, except for the narrow $\unit[0.25]{mm}$-wide sheath region in front of the momentary cathode  
where $\mathrm{Ar}_2^+$ ions have a higher density than the electrons. 
As in the positive half-period, $\mathrm{Ar}^+$  ions are effectively converted to $\mathrm{Ar}_2^+$ by atomic-to-molecular ion conversion. 
The dominant charge carrier loss process is dissociative recombination, as shown in \fref{RatesNHP}(a).

With the increase of the absolute value of the gap voltage, the electric field in the sheath region increases (\fref{RatesNHP}(a)). 
This enhances the influx of ions to the momentary cathode and supports electron multiplication during the Townsend pre-phase. 
It also leads to an increase of the mean electron energy in the sheath region. 
The mean electron energy reaches a value of $\unit[7]{eV}$ near the momentary cathode, while its remains lower than $\unit[1]{eV}$ in the remainder of the gap (\fref{DischargeNHP}(a)).
During the further development, the secondary electrons gain enough energy to excite and ionise the ground state atoms leading to the accumulation of a positive space charge and consequently the further increase of the electric field near the momentary cathode. 
When the critical value of the space charge, i.e.\ the electric field, is reached, the positive streamer onset occurs.

Upon reaching the momentary cathode, a thin cathode layer is formed, where the reduced electric field reaches a value of more than $\unit[4600]{Td}$.  
The mean electron energy is highest in the sheath. 
It ranges between $\unit[8]{eV}$ just at the momentary cathode and $\unit[6]{eV}$ approx.~$\unit[0.5]{mm}$ away from it. Then it decays towards the momentary anode from $\unit[6]{eV}$ to less than $\unit[1]{eV}$. 
Due to the maximum of the mean electron energy in the thin sheath layer, a strong production of electrons, ions and excited states takes place just in front of the momentary cathode~(\fref{DischargeNHP}(b) and \fref{RatesNHP}(b)). 
The electron density reaches a value of~$\unit[8 \times 10^{20}]{m^{-3}}$, which is almost three orders of magnitude higher than in the rest of the discharge channel. 
This value is one order of magnitude higher than the one observed during the streamer phase in the positive half-period, which can be explained by the fact that the streamer approaches a metal surface where no charges are accumulated.
Behind the streamer head, the increase of the ionisation of excimers and excited atoms can be observed in~\fref{RatesNHP}(b). 
These processes gradually lead to the start of the transient glow  phase.

During the transient glow  phase, the ionisation wave moves towards the momentary cathode 
and produces charge carriers all along the discharge channel. This is also illustrated by the increase of the electron density along the gap represented in \fref{2Dt-NHP}. 
Simultaneously, the density of most of the species increases by almost one order of magnitude in the entire gap (\fref{DischargeNHP}(c)).
For instance, the density of electrons and $\mathrm{Ar}_2^+$ ions reaches a value of about~$\unit[2 \times 10^{21}]{m^{-3}}$ in front of the momentary cathode, while in the rest of the gap it is above~$\unit[10^{20}]{m^{-3}}$. 
The mean electron energy also increases throughout the gap, reaching an average value of about~$\unit[4]{eV}$ (\fref{DischargeNHP}(c)). However, note that the reduced electric field in the whole gap remains quite low with a value of less than~$\unit[3]{Td}$ (cf.~\fref{DischargeNHP}(c)). 
Due to this, electrons cannot gain enough energy for direct impact ionisation. 
The dominant processes for electron production are now solely the stepwise ionisation near the momentary cathode and the ionisation of excimers in the rest of the gap (\fref{RatesNHP}(c)),  chemo-ionisation processes are orders of magnitude lower. 
The local maxima of argon excimers now lead to the localised increase of ionisation.
About~$\unit[20]{ns}$ after the arrival of the streamer at the cathode, striated structures emerge near the momentary anode~(\fref{RatesNHP}(c)). 
Similar as in the positive half-period, the localised space charge surplus causes localised maxima and minima in the reduced electric field varying between $\unit[0]{}$ and $\unit[5]{Td}$ and enhancing the striations.
It is important to note that the striations appear at the same position as in the positive half-period discharge. 
This means that the location of the striations is not related to the momentary cathode but to the position of the dielectric and the location of the excited species maxima.

In general, the discharge in the negative half-period is similar to the discharge in the positive half-period. 
The main difference between the two discharge events is the absence of the surface discharge leading to a lower discharge current and a faster transition from the streamer phase to the transient glow  phase during the negative half-period. 
Note that the stronger electron production during the streamer phase also contributes to the faster transition between the two phases. 
Based on this, it follows that the surface discharge and consequently the surge of secondary electrons originating from the impingement of positive ions on the surface is responsible for the strong peaks of the electron density and discharge current and the appearance of pronounced striations during the discharge in the positive half-period. 
Although the charge carrier densities in the plasma bulk are much higher during the positive half-period, 
it is interesting that the spatial modulation of the electric field and mean electron energy remains at similar order of magnitude in both half-periods. 

In conclusion, the mechanism of the striation formation is similar in both half-periods. 
The striations are caused by the appearance of an ionising wave in the gap, induced by secondary electrons. 

\section{Summary}
\label{Cnclsn}
The mechanisms of striation formation in a sine-driven single-filament atmospheric-pressure DBD in argon were investigated using fluid modelling. 
A detailed analysis of positive and negative half-periods in the quasi-periodic state was performed to investigate the influence of volume and surface memory effects.
Striated structures along the discharge channel were observed during the discharges in both the positive and negative half-period. 
This confirms that the stratification of filamentary DBDs can be analysed by the used fluid-Poisson model. 
The provided high spatial and temporal resolution of the model allows the identification of the main mechanisms responsible for striated structure formation.

It was found that the formation of striations occurs during the transient glow phase of the DBD, which follows the Townsend pre-phase and the streamer phase in both half-periods.
Compared to the discharge during the positive half-period, the discharge during the negative half-period exhibited a faster transition from the streamer to the glow  phase, weaker striations, and a longer discharge duration.
Moreover, it was found that the striations in positive and negative half-periods appear at the same position, i.e.\ the location of the striations is not related to the momentary cathode but to the position of the dielectric.

The striated structure formation was explained by the disturbance of the spatial distribution of the electrons along the discharge channel due to repetitive ionisation waves. 
The main contributions to the electron production were found to be stepwise ionisation of metastable argon atoms and ionisation of excimers. 
The spatial distribution of the excited atoms and molecules, thus, directly determines the location of the local maxima of the electron production.
Excessive excitation was found in the sheath region during the streamer phase. 
The secondary electrons emitted during the surface discharge ionise these excited atoms and molecules in the form of an ionisation wave. 
In turn, space charges form near the excited state maxima, leading to a modulation of the electric field and further enhancement of the ionisation and striation formation.

\ack
This work was funded by the Deutsche Forschungsgemeinschaft (DFG, German Research Foundation)---project numbers 407462159 and 466331904,  the Czech Science Foundation---project No. 21-16391S and supported by project CEPLANT (LM2023039) funded by the Ministry of Education, Youth and Sports of the Czech Republic.

\pagebreak 
\clearpage

\appendix
\onecolumn
\section*{Appendix A. Reaction kinetic processes}

\begin{table*}[htp!]
\begin{threeparttable}
\caption{Electron collision processes considered in the reaction kinetics model for argon. The rate coefficients are expressed in $\mathrm{m}^3/\mathrm{s}$ and the range of kinetic electron energy is given in $\mathrm{eV}$.}
\begin{indented}
\item[]\begin{tabular}{@{}l*{15}{l}||}
\br
Index & Process & Energy range & Data source\\
\mr
\multicolumn{2}{l}{\textit{Elastic electron collision}} & & \\
1 & $\mathrm{Ar} + \mathrm{e} \rightarrow \mathrm{Ar}+ \mathrm{e}$ & $0 - 1000$ & \cite{Hayashi2003} \\
\multicolumn{2}{l}{\textit{Electron-collision excitation and deexcitation}} & & \\
2 & $\mathrm{Ar} + \mathrm{e}   \rightarrow \mathrm{Ar}^* + \mathrm{e} $ & $11.55 - 1000$ & \cite{Hayashi2003} \\
3 & $\mathrm{Ar}^* + \mathrm{e}     \rightarrow \mathrm{Ar} + \mathrm{e} $  & $0-1000$  \\
\multicolumn{2}{l}{\textit{Electron-collision ionisation}} & & \\
4 & $\mathrm{Ar} + \mathrm{e} \rightarrow \mathrm{Ar}^+ + 2\mathrm{e} $  & $15.76 - 1000$ & \cite{Rapp-1965-ID5358}\\
5 & $\mathrm{Ar}^* + \mathrm{e}     \rightarrow \mathrm{Ar}^+ + 2\mathrm{e} $ &  $\geq 4.21$& \cite{Vriens-1980-ID365} \\
6 & $\mathrm{Ar}_2^* + \mathrm{e}    \rightarrow \mathrm{Ar}_2^+ + 2\mathrm{e}$ & $3.23-50; \geq 50$ & \cite{Flannery1980}; Born approximation \\
\br
\end{tabular}
\begin{tablenotes}
\small
\item[a)] The deexcitation is taken into account in the reaction scheme by principle of detailed balance where cross section for this process is obtained as:
\begin{equation*}
Q^\mathrm{deexc}(u_\mathrm{e}) = \frac{g_m}{g_m^*} \frac{u_\mathrm{e}-u_\mathrm{exc}}{u_\mathrm{e}}Q^\mathrm{exc}(u_\mathrm{e}+u_\mathrm{exc})
\label{DB}
\end{equation*}
where $g_m=1$  and $g_m^* = 48$ denote the statistical weights of a given level and $Q^\mathrm{exc}(u_\mathrm{e}+u_\mathrm{exc})$ is excitation cross section (taken from \cite{Hayashi2003}).
\end{tablenotes}
\end{indented}
\end{threeparttable}
\end{table*}

\begin{table*}[h!]
\begin{threeparttable}
\caption{Heavy particle collision and radiative processes considered in the reaction kinetics model for argon. The rate coefficients of two-body collisions are expressed in $\mathrm{m}^3/\mathrm{s}$, those of three-body collisions are in $\mathrm{m}^6/\mathrm{s}$ and the coefficient for radiation processes are in units of $1/\mathrm{s}$. The temperatures $T_\mathrm{e}$ and $T_\mathrm{g}$ are given in $\mathrm{K}$.}
\begin{indented}
\item[]\begin{tabular}{@{}l*{15}{l}||}
\br
Index & Process & Rate coefficient & Data source\\
\mr
\multicolumn{2}{l}{\textit{Electron-ion recombination}} & & \\
7 & $\mathrm{Ar}^+ + 2\mathrm{e} \rightarrow \mathrm{Ar}^*  + \mathrm{e} $ & $1.0 \times 10^{-31} (T_\mathrm{e}/300)^{-4.5}$ \tnote{a)} & \cite{Eletskii-1986-ID3349} \\
8 & $\mathrm{Ar}_2^+ + \mathrm{e} \rightarrow \mathrm{Ar}^* + \mathrm{Ar}$ & $6.86 \times 10^{-13}(T_\mathrm{e}/300)^{-2/3} (1.0-\mathrm{e}^{-418/T_\mathrm{g}})/(1.0-0.31 \mathrm{e}^{-418/T_\mathrm{g}})$ & \cite{Cunningham-1981-ID3076} \\
\multicolumn{2}{l}{\textit{Chemoionisation process}} & & \\
9 & $\mathrm{Ar}^* + \mathrm{Ar}^*  \rightarrow \mathrm{Ar}^+ + \mathrm{Ar} + \mathrm{e} $  & $1.20 \times 10^{-15}$ \tnote{b)} &  \cite{Dyatko-2008-ID2506} \\
\multicolumn{2}{l}{\textit{Neutral association process}} & & \\
10 & $\mathrm{Ar}^* + 2\mathrm{Ar} \rightarrow \mathrm{Ar}_2^* + \mathrm{Ar}$  & $8.33 \times 10^{-45}$ & \cite{Eletskii-1986-ID3349} \\
\multicolumn{2}{l}{\textit{Charge-transfer process}} & & \\
11 & $\mathrm{Ar}^+ + 2\mathrm{Ar} \rightarrow \mathrm{Ar}_2^+ + \mathrm{Ar}$  &  $2.25 \times 10^{-43}(T_\mathrm{g}/300)^{-0.4}$ & \cite{Jones-1980-ID3414} \\
\multicolumn{2}{l}{\textit{Radiative processes}} & & \\
12 & $\mathrm{Ar}^* \rightarrow \mathrm{Ar} + h\nu$ & $6.70 \times 10^5$ \tnote{c)}  &  \\
13 & $\mathrm{Ar}_2^* \rightarrow 2\mathrm{Ar} + h\nu$ & $3.13\times10^5$ & \cite{Sigeneger-2016-ID3928} \\
\br
\end{tabular}
\begin{tablenotes}
\small
\item[a)] The remaining electron gains all of the excess of the energy.
\item[b)] Same as in reaction 7, the electron gains all of the excess of the energy.
\item[c)] The radiation probability is calculated according to the Holstein's theory for a cylindrical plasma as $k_\mathrm{eff} = 0.196 k_0 \sqrt{\lambda_0/R_\mathrm{f}}$~\cite{Sigeneger-2016-ID3928, Holstein-1951-ID53}, with approximate filament radius $R_\mathrm{f}$ =  $10^{-4} \, \unit[]{m}$, wavelength $\lambda_0 = 0.106 \times 10^{-6} \,\unit[]{m}$ and transition probability (between 1s$_4$ and the ground state) $k_0$ = $1.056 \times 10^{8} \, \unit[]{s^{-1}}$ taken from \cite{Irimia-2004-ID2181}.
\end{tablenotes}
\end{indented}
\end{threeparttable}
\end{table*}

\pagebreak 
\clearpage

\twocolumn
\section*{References}
\bibliographystyle{iopart-num}
\bibliography{mybib}

\end{document}